\documentclass[letterpaper,12pt]{article}

\usepackage{amsmath}

\usepackage{mathptmx}
\usepackage{graphics}

\usepackage[authoryear,comma,longnamesfirst,sectionbib]{natbib}

\usepackage[dvips]{graphicx}
\usepackage[colorlinks=true, hypertex, linkcolor=blue, urlcolor=blue, citecolor=blue]{hyperref}
\usepackage{booktabs} 
\numberwithin{figure}{subsection}
\numberwithin{table}{subsection}

\title{A Regression-based Adjusted Plus-Minus Statistic for NHL Players}
\author{Brian Macdonald}

\begin{document}
\maketitle
\begin{abstract}
The goal of this paper is to develop an adjusted plus-minus statistic for NHL players that is independent of both teammates and opponents.  We use data from the shift reports on NHL.com in a weighted least squares regression to estimate an NHL player's effect on his team's success in scoring and preventing goals at even strength.  Both offensive and defensive components of adjusted plus-minus are given, estimates in terms of goals per 60 minutes and goals per season are given, and estimates for forwards, defensemen, and goalies are given.
\end{abstract}

{\bf Keywords:} plus-minus, hockey, nhl, sports
\newpage
\tableofcontents
\newpage
\listoftables
\listoffigures
\newpage
\section{Introduction}\label{intro}
    Hockey analysts have developed several metrics that attempt to quantify an NHL player's contribution to his team.  Tom Awad's Goals Versus Threshold in \cite{awad}, Jim Corsi's Corsi rating as described in \cite{boersma}, Gabriel Desjardins' Behindthenet Rating, along with his on-ice/off-ice, strength of opponents, and strength of linemates statistics in \cite{gabe}, Iian Fyffe's Point Allocation in \cite{fyffe}, Ken Krzywicki's shot quality, as presented in \cite{ken1} and updated in \cite{ken2},  Alan Ryder's Player Contribution in \cite{ryder}, and Timo Seppa's Even-Strength Total Rating in \cite{seppa} are a few examples.  In this paper, we propose a new metric, adjusted plus-minus ($APM$), that attempts to estimate a player's contribution to his team in even-strength (5-on-5, 4-on-4, and 3-on-3) situations, independent of that player's teammates and opponents, and in the units of goals per season.  $APM$ can also be expressed in terms of goals per 60 minutes ($APM/60$).  We find both an adjusted offensive plus-minus component ($OPM$) and an adjusted defensive plus-minus component ($DPM$), which estimate the offensive and defensive production of a player at even-strength, independent of teammates and opponents, and in the units of goals per season.

    Inspired by the work in basketball by \cite{rosenbaum}, \cite{ilardibarzilai}, and \cite{eli}, we use weighted multiple linear regression models to estimate $OPM$ per 60 minutes ($OPM/60$) and $DPM$ per 60 minutes ($DPM/60$).  The estimates are a measure of the offensive and defensive production of a player in the units of goals per 60 minutes.  These statistics, along with average minutes played per season, give us $OPM$ and $DPM$.  Adding $OPM/60$ and $DPM/60$ gives $APM/60$, and adding $OPM$ and $DPM$ gives $APM$.  We emphasize that we consider only even-strength situations.

    The main benefit of the weighted linear regression model is that the resulting adjusted plus-minus statistics for each player should in theory be independent of that player's teammates and opponents.  The traditional plus-minus statistic in hockey is highly dependent on a player's teammates and opponents, and the use of the regression removes this dependence.  One drawback of our model is statistical noise.  In order to improve the estimates and reduce the standard errors in the estimates, we use data from three NHL seasons, and we combine the results of two different models, one inspired by \cite{ilardibarzilai}, and the other by \cite{rosenbaum}.

\subsection{Example of the Results}
Before we describe the models in detail, we give the reader an example of the results.  The typical NHL fan has some idea of who the best offensive players in the league are, so we give the top 10 players in average $OPM$ during the 2007-08, 2008-09, and 2009-10 seasons, sorted by $OPM$, in Table \ref{opmp}.
\begin{table}[h!]
\begin{center}
\caption{
Top 10 Players in OPM
}
\label{opmp}
{\small
\begin{tabular}{llrrrrrrrrr}
  \addlinespace[.3em] \toprule
Rk & Player & Pos & OPM & OErr & DPM & APM & Mins & GF60 & OPM60 & GF \\
  \midrule
  1 & Pavel Datsyuk & C & 15.4 & 3.4 & 6.2 & 21.6 & 1186 & 3.39 & 0.777 &  67 \\
    2 & Alex Ovechkin & LW & 15.2 & 3.8 & 0.2 & 15.4 & 1262 & 3.69 & 0.723 &  78 \\
    3 & Sidney Crosby & C & 14.4 & 2.6 & $-$0.9 & 13.5 & 1059 & 3.59 & 0.818 &  63 \\
    4 & Henrik Sedin & C & 14.0 & 4.5 & $-$5.7 & 8.3 & 1169 & 3.35 & 0.718 &  65 \\
    5 & Evgeni Malkin & C & 13.2 & 2.7 & $-$3.0 & 10.2 & 1164 & 3.37 & 0.681 &  65 \\
    6 & Zach Parise & LW & 12.6 & 3.2 & 3.0 & 15.6 & 1164 & 2.94 & 0.652 &  57 \\
    7 & Joe Thornton & C & 12.0 & 3.2 & 3.6 & 15.6 & 1222 & 3.13 & 0.590 &  64 \\
    8 & Eric Staal & C & 11.5 & 3.1 & $-$0.5 & 11.0 & 1159 & 3.00 & 0.594 &  58 \\
    9 & Ilya Kovalchuk & LW & 10.9 & 2.8 & $-$4.2 & 6.7 & 1189 & 2.93 & 0.551 &  58 \\
   10 & Marian Gaborik & RW & 10.2 & 2.2 & 4.3 & 14.5 & 853 & 3.28 & 0.715 &  47 \\
   \bottomrule
\end{tabular}
}
\end{center}
\end{table}
Note that $Rk$ is the rank of that player in terms of $OPM$, $Pos$ is the player's position, $OErr$ is the standard error in the $OPM$ estimates, $Mins$ is the number of minutes that the player played on average during the 2007-08, 2008-09, and 2009-10 seasons, $GF60$ are the goals per 60 minutes that a player's team scored while he was the ice at even-strength, and $GF$ are the goals per season that a player's team scored while he was the ice at even-strength.  The 10 players in this list are arguably the best offensive players in the game.  Ovechkin, Crosby, Datsyuk, and Malkin, perhaps the league's most recognizable superstars, make the top 5 along with Henrik Sedin, who led the NHL in even-strength points during the 2009-2010 season with 83 points, which was 10 more points than the next leading scorer.

We highlight two interesting numbers in this list.  First, note that Pavel Datsyuk, who is regarded by many as the best two-way player in the game, has the highest defensive rating among these top offensive players.  Datsyuk's excellent two-way play gives him the highest $APM$ estimate among forwards and defensemen.  We give the list of top 10 forwards and defensemen in $APM$, and discuss several other top 10 lists for $OPM/60, OPM, DPM/60, DPM, APM/60,$ and $APM$, in Section \ref{resultssummary}.  Second, note that Henrik Sedin has a much higher $OErr$ than the other players in the list.  This increased error is likely due to the fact that Henrik plays most of his minutes with his brother Daniel, and the model has a difficult time separating the contributions of the twin brothers.  The Sedin twins provide us with a great example to use when analyzing the errors, and we discuss the Sedins and the errors in more detail in Section \ref{errors}.

\subsection{Complete Results}
A \verb .csv  file containing the complete results can be obtained by contacting the author.  An interested reader may prefer to open these results in a spreadsheet program and filter by position or sort by a particular statistic.  Also, the \verb .csv  file contains more columns than the list given in Table \ref{opmp}.  For example, the file includes the three most frequent linemates of each player along with the percentage of minutes played with each of those linemates during the 2007-08, 2008-09, and 2009-10 NHL regular seasons.  An example of these additional columns is given in Table \ref{linemateexample}.
\begin{table}[h!]
\begin{center}
\caption{Example of Linemate Details}
\label{linemateexample}
{\small
\begin{tabular}{lrrrrrrrr}
  \addlinespace[.3em] \toprule
Player & Pos & Mins & Teammate.1 & min1 & Teammate.2 & min2 & Teammate.3 & min3 \\
  \midrule
Henrik Sedin & C & 1169 & D.Sedin & 83\% & R.Luongo & 76\% & A.Edler & 35\% \\
  Daniel Sedin & LW & 1057 & H.Sedin & 92\% & R.Luongo & 77\% & A.Edler & 35\% \\
   \bottomrule
\end{tabular}
}
\end{center}
\end{table}
\noindent Notice that, as suggested in the table, Henrik Sedin played 83\% of his minutes with brother Daniel, and Daniel played 92\% of his minutes with Henrik.

Finally, the file includes columns for the goals a player's team scored ($GF$), the goals a player's team allowed ($GA$), and the net goals a player's team scored ($NG$), while he was on the ice at even-strength.  These statistics are in terms of average goals per season during the 2007-08, 2008-09, and 2009-10 NHL regular seasons.  We also give $GF/60, GA/60,$ and $NG/60$, which are $GF, GA,$ and $NG$ in terms of goals per 60 minutes.  An example of this information is given in Table \ref{goalsexample}.
\begin{table}[h!]
\begin{center}
\caption{Example of GF, GA, and NG statistics}
\label{goalsexample}
{\small
\begin{tabular}{lrrrrrrr}
  \addlinespace[.3em] \toprule
Player & Pos & GF60 & GA60 & NG60 & GF & GA & NG \\
  \midrule
Sidney Crosby & C & 3.59 & 2.55 & 1.04 &  63 &  45 &  18 \\
  Pavel Datsyuk & C & 3.39 & 1.84 & 1.55 &  67 &  36 &  31 \\
   \bottomrule
\end{tabular}
}
\end{center}
\end{table}
These raw statistics, along with the linemate information, will be helpful in the analysis of the results of our model.

The rest of this paper is organized as follows.  In Section \ref{models}, we describe the two models we use to compute $OPM$, $DPM$, and $APM$.  In Section \ref{resultssummary}, we summarize and discuss the results of these models by giving various top 10 lists, indicating the best forwards, defensemen, and goalies according to $OPM$, $DPM$, and $APM$, as well as their corresponding per 60 minute statistics.  Section \ref{discussion} contains a discussion of the model.  We summarize and discuss the advantages \eqref{adv} and disadvantages \eqref{disadv} of these statistics.  Next, we give more details about the formation of the model, including the selection of the variables (\ref{variables}), and selection of the observations (\ref{observations}).  Also, we discuss our assumptions (\ref{assumptions}) as well as the standard errors (\ref{errors}) in the estimates.  We finish with ideas for future work and some conclusions \eqref{futurework}.

\section{Two weighted least-squares models}\label{models}
    We now define our variables and state our models.  In each model, we use players who have played a minimum of 4000 shifts over the course of the 2007-08, 2008-09, and 2009-10 seasons  (see Section \ref{discussion} for a discussion).  We define a shift to be a period of time during the game when no substitutions are made.  The observations in each model are weighted by the duration of that observation in seconds.

    \subsection{Ilardi-Barzilai-type model}\label{model1}
    Inspired by \cite{ilardibarzilai}, we use the following linear model:
        \begin{align}\label{model1eq}
            y = \beta_0 + \beta_1 X_1 + \cdots + \beta_J X_J &+ \delta_1 D_1 + \cdots + \delta_J D_J + \gamma_1 G_1 + \cdots + \gamma_K G_K + \epsilon,
        \end{align}
    \noindent where $J$ is the number of skaters in the league, and $K$ is the number of goalies in the league. The variables in the model are defined as follows:
    \begin{align*}
        y   &= \text{goals per 60 minutes during an observation} \\
        X_j &= \left\{
                \begin{array}{ll}
                  1, & \hbox{ skater $j$ is on offense during the observation;} \\
                  0, & \hbox{ skater $j$ is not playing or is on defense during the observation;}
                \end{array}
              \right.
        \\
        D_j &= \left\{
                \begin{array}{ll}
                  1, & \hbox{ skater $j$ is on defense during the observation;} \\
                  0, & \hbox{ skater $j$ is not playing or is on offense during the observation;}
                \end{array}
              \right.
        \\
        G_k &= \left\{
                \begin{array}{ll}
                  1, & \hbox{ goalie $k$ is on defense during the observation;} \\
                  0, & \hbox{ goalie $k$ is not playing or is on offense during the observation;}
                \end{array}
              \right.
    \end{align*}
    where $1 \leq j \leq J,$ and $1\leq k \leq K$.  Note that by ``skater" we mean a forward or a defensemen, but not a goalie.  The coefficients in the model have the following interpretation:
    \begin{align}\label{coeffs}
         \beta_j     &= \text{goals per 60 minutes contributed by skater $j$ on offense,} \notag\\
        -\delta_j    &= \text{goals per 60 minutes contributed by skater $j$ on defense,} \notag\\
        -\gamma_k    &= \text{goals per 60 minutes contributed by goalie $k$ on defense,}  \\
         \beta_0     &= \text{intercept,} \notag\\
         \epsilon    &= \text{error.}\notag
    \end{align}
    The coefficient $\beta_1$, for example, gives an estimate, in goals per 60 minutes, of how $y$ changes when Skater 1 is on the ice on offense ($X_1=1$) versus when Skater 1 is not on the ice on offense ($X_1=0$), independent of all other players on the ice.  The coefficients $\beta_j, -\delta_j,$ and $-\gamma_k$ are estimates of $OPM/60$ for Skater $j$, $DPM/60$ for Skater $j$, and $DPM/60$ for Goalie $k$, respectively.  They are playing-time-independent rate statistics, measuring the offensive and defensive value of a player in goals per 60 minutes.

    Notice the negative sign in front of $\delta_j$ and $\gamma_k$ in \eqref{coeffs}.  Note that a negative value for one of these coefficients corresponds to a positive contribution.  For example, if Skater 1 has a defensive coefficient of $\delta_1 = -0.8$, he prevents $0.8$ goals per 60 minutes when he is on defense.
    We could have chosen to define a skater's $DPM/60$ to be $+\delta_j$, in which case negative values for $DPM/60$ would be good. Instead, we prefer that positive contributions be represented by a positive number, so we define $DPM/60 = -\delta_j$ for skaters.  Likewise, we define $DPM/60 = -\gamma_k$ for goalies.  For Skater 1's $DPM/60$ in our example, we have
        $$ DPM/60 = -\delta_1 = -(-0.8) = +0.8, $$
    which means that Player 1 has a positive contribution of +0.8 goals per 60 minutes on defense.

    Note that for the observations in this model, each shift is split into two lines of data: one line corresponding to the home team being on offense, and one line corresponding to the away team being on offense.  It is assumed that in hockey, unlike in other sports, a team plays offense and defense concurrently, and the two observations for each shift are given equal weight.    Also, note that we include separate defensive variables for goalies, but no offensive variables.  Here we are assuming that goalies do not contribute on offense.  See Section \ref{assumptions} for a discussion of these assumptions.
    Finally, we note that the data used for the model were obtained from the shift charts published in \cite{nhlcom} for games played in the 2007-08, 2008-09, and 2009-10 regular seasons.  See Section \ref{observations} for more about the data used and to see how it was selected.

    \subsection{Calculating $OPM$, $DPM$, and $APM$}\label{apm}

    A player's contribution in terms of goals over an entire season is useful as well, and may be preferred by some NHL fans and analysts.  We use the regression coefficients and minutes played to give playing-time-dependent counting statistic versions of the rate statistics from the regression model.  These counting statistics are $OPM$, $DPM$, and $APM$, and they measure the offensive, defensive, and total value of a player, in goals per season.  To get a skater's $OPM$, for example, we multiply a skater's offensive contribution per minute by the average number of minutes that the skater played per season from 2007-2010.  The value for $DPM$ is found likewise, and $APM$ for a player is the sum of his $OPM$ and his $DPM$.  Goalies have no $OPM$, so a goalie's $APM$ is simply his $DPM$.  Let
    \begin{align*}
        MinO_j &= \text{minutes per season on offense for skater $j$,}\\
        MinD_j &= \text{minutes per season on defense for skater $j$, and}\\
        MinG_k &= \text{minutes per season on defense for goalie $k$.}
    \end{align*}
    Then, we can calculate $OPM, DPM$, and $APM$ for skaters and goalies as follows:
    \begin{align}\label{apmformulas}
        \hskip 1.25in OPM_j &= \,\,\,\,\,\,\, \beta_j \,  MinO_j/60,                           \notag\\
        \hskip 1.25in DPM_j &= -\delta_j \, MinD_j/60  &\text{ (for skaters),   \hskip 1in }\notag\\
        \hskip 1.25in DPM_k &= -\gamma_k \, MinG_k/60  &\text{ (for goalies),   \hskip 1in }     \\
        \hskip 1.25in APM_j &= OPM_j + DPM_j          &\text{ (for skaters),   \hskip 1in }\notag\\
        \hskip 1.25in APM_k &= DPM_k                  &\text{ (for goalies).   \hskip 1in }\notag
    \end{align}
    In order to estimate $Err$, the standard errors for the $APM$ estimates, we assume that $OPM$ and $DPM$ are uncorrelated, and we have
        \begin{align*}
            {Err} = \sqrt{Var(APM)} = \sqrt{({OErr})^2 + ({DErr})^2}.
        \end{align*}
        where $OErr$ and $DErr$ are the standard errors in the $OPM$ and $DPM$ estimates, respectively, and $Var$ is variance.  The assumption that offensive and defensive contributions of a player are uncorrelated is debatable. See, for example, \cite{corey}.

\subsection{Rosenbaum-type model}\label{model2}
    In an effort to improve the estimates and their errors, we use a second linear model, this one inspired by \cite{rosenbaum}:
    \begin{align}\label{model2eq}
    y_{net} = \eta_0 + \eta_1 N_1 + \cdots + \eta_{J+K} N_{J+K} + \epsilon.
    \end{align}
     The variables in the model are defined as follows:
    \begin{align*}
        y_{net}   &= \text{net goals per 60 minutes for the home team during an observation} \\
        N_j &= \left\{
                \begin{array}{ll}
                  \,\,\,\,\,\,\, 1, & \hbox{ player $j$ is on the home team during the observation;} \\
                 -1, & \hbox{ player $j$ is on the away team during the observation;} \\
                  \,\,\,\,\,\,\, 0, & \hbox{ player $j$ is not playing during the observation.}
                \end{array}
              \right.
        \end{align*}
    The coefficients in the model have the following interpretation:
    \begin{align*}
        \eta_j     &= \text{net goals per 60 minutes contributed by player $j,$} \\
        \eta_0     &= \text{intercept}, \\
        \epsilon    &= \text{error}.
    \end{align*}
    The coefficients $\eta_1, \ldots, \eta_J$ of this model are estimates of each player's $APM/60$.
By ``net goals" we mean the home team's Goals For ($GF$) minus the home team's Goals Against ($GA$).  Also, note that by ``player" we mean forward, defensemen, or goalie.
    The data used for these variables were also obtained from the shift charts published on NHL.com for games played in the 2007-08, 2008-09, and 2009-10 regular seasons.  
    In this model, unlike the Ilardi-Barzilai model, each observation in this model is simply one shift.  We do not split each shift into two lines of data.

    In order to separate offense and defense, we follow Rosenbaum and form a second model:
    \begin{align}\label{model3eq}
    y_{tot} = \tau_0 + \tau_1 T_1 + \cdots + \tau_{J+K} T_{J+K} + \epsilon.
    \end{align}
    The variables in the model are defined as follows:
        \begin{align*}
            y_{tot}   &= \text{total goals per 60 minutes scored by both teams during an observation} \\
            T_j &= \left\{
                    \begin{array}{ll}
                      1, & \hbox{ player $j$ is on the ice (home or away) during the observation;} \\
                      0, & \hbox{ player $j$ is not on the ice during the observation.}
                    \end{array}
                  \right.
        \end{align*}
    The coefficients in the model have the following interpretation:
        \begin{align*}
            \tau_j     &= \text{total goals per 60 minutes contributed by skater $j$}, \\
            \tau_0     &= \text{intercept}, \\
            \epsilon    &= \text{error}.
        \end{align*}
    By total goals, we mean $GF+GA$.  Recall that the coefficients in \eqref{model2eq} were estimates of each player's $APM/60$, or net goals contributed per 60 minutes.  Likewise, the coefficients in \eqref{model3eq} are estimates of each player's $TPM/60$, or total goals contributed per 60 minutes.
    In \eqref{apmformulas}, we used playing time to convert $APM/60$ to $APM$, and likewise, we can convert $TPM/60$ to $TPM.$   
    
    We know from before that
    \begin{align}\label{apmformula}
        APM/60 = OPM/60 + DPM/60,
    \end{align}
    and we also have that
    \begin{align} \label{tpmformula}
        {TPM/60} = {OPM/60 } - { DPM/60}.
    \end{align}
    Using equations \eqref{apmformula} and \eqref{tpmformula}, if we add a player's $TPM/60$ and $APM/60$, and divide by 2, the result is that player's $OPM/60$:
    \begin{align*}
         &\frac{1}{2}(APM/60 + TPM/60)  
         =  OPM/60.
        \end{align*}
    Likewise,
    \begin{align*}
        \frac{1}{2} (APM/60 - TPM/60) = DPM/60.
    \end{align*}
Using playing time, we can convert $OPM/60$, $DPM/60$, and $APM/60$ to $OPM$, $DPM$, and $APM$ as we did with our first model in Section \ref{model1}.  Note that in this model, unlike the model in Section \ref{model1}, all players are treated the same, which means that the model gives offensive estimates for goalies and skaters alike.  While a goalie can impact a team's offensive production, we typically do not use these offensive estimates for goalies.  Goalies and offense are discussed more in Section \ref{goalies}. 

\subsection{Averaging results from the two models}
    The estimates obtained from the models in Section \ref{model1} and \ref{model2} can be averaged, and the resulting estimates will have smaller standard errors than the individual estimates from either of the two models.  Let $OPM^{ib}_j$, $DPM^{ib}_j$, and $APM^{ib}_j$ be the $OPM$, $DPM$, and $APM$ results for player $j$ from the Ilardi-Barzilai-type model (Section \ref{model1}), and likewise let $OPM^r_j$, $DPM^r_j$, and $APM^r_j$ be the corresponding results from the Rosenbaum-type model (Section \ref{model2}).  We average the results from our two models to arrive at our final metrics $OPM$, $DPM$, and $APM$:
    \begin{align*}
        OPM_j &= \frac{1}{2} (OPM^{ib}_j + OPM^r_j),  \\
        DPM_j &= \frac{1}{2} (DPM^{ib}_j + DPM^r_j),  \\
        APM_j &= \frac{1}{2} (APM^{ib}_j + APM^r_j).
    \end{align*}
Each model has its advantages and disadvantages, so we have chosen to weight the results from the two models equally.  See Section \ref{goalies} for a discussion of the benefits and drawbacks of each model.  Assuming the errors are uncorrelated we can estimate them as follows:
    \begin{align*}
        OErr_j &= \frac{1}{2} \sqrt{(OErr^{ib}_j)^2 + (OErr^r_j)^2 }, \\
        DErr_j &= \frac{1}{2} \sqrt{(DErr^{ib}_j)^2 + (DErr^r_j)^2 }, \\
         Err_j &= \frac{1}{2} \sqrt{( Err^{ib}_j)^2 + ( Err^r_j)^2 }.
    \end{align*}
Note that the errors $OErr_j$ are smaller than the errors $OErr_j^{r}$ and $OErr_j^{ib}$.  Likewise, $DErr_j$ and $Err_j$ are smaller than each of the components used to compute them.

\section{Summary of Results}\label{resultssummary}
    In this section we will summarize the results of the model by giving various top 10 lists, indicating the best offensive, defensive, and overall players in the league according to the estimates found in the model.
    \subsection{$OPM/60$}  Recall that $OPM/60$ is a measure of the offensive contribution of a player at even-strength in terms of goals per 60 minutes of playing time.  Recall also that we assume that goalies do not contributed on offense, so we list only forwards and defensemen in this section.
\begin{table}[h!]
\begin{center}
\caption{
Top 10 Players in OPM60
}
\label{opm60p}
{\small
\begin{tabular}{llrrrrrrrrr}
  \addlinespace[.3em] \toprule
Rk & Player & Pos & OPM60 & OErr & DPM60 & APM60 & Mins & GF60 & OPM & GF \\
  \midrule
  1 & Sidney Crosby & C & 0.818 & 0.148 & $-$0.052 & 0.766 & 1059 & 3.59 & 14.4 &  63 \\
    2 & Pavel Datsyuk & C & 0.777 & 0.174 & 0.314 & 1.091 & 1186 & 3.39 & 15.4 &  67 \\
    3 & Alex Radulov & RW & 0.758 & 0.222 & $-$0.248 & 0.510 & 343 & 3.67 & 4.3 &  21 \\
    4 & Alex Ovechkin & LW & 0.723 & 0.178 & 0.010 & 0.733 & 1262 & 3.69 & 15.2 &  78 \\
    5 & Henrik Sedin & C & 0.718 & 0.231 & $-$0.294 & 0.424 & 1169 & 3.35 & 14.0 &  65 \\
    6 & Marian Gaborik & RW & 0.715 & 0.155 & 0.303 & 1.018 & 853 & 3.28 & 10.2 &  47 \\
    7 & Evgeni Malkin & C & 0.681 & 0.141 & $-$0.156 & 0.525 & 1164 & 3.37 & 13.2 &  65 \\
    8 & Zach Parise & LW & 0.652 & 0.166 & 0.155 & 0.807 & 1164 & 2.94 & 12.6 &  57 \\
    9 & Jakub Voracek & RW & 0.642 & 0.186 & $-$0.045 & 0.597 & 621 & 3.03 & 6.6 &  31 \\
   10 & C. Gunnarsson & D & 0.608 & 0.245 & 0.233 & 0.841 & 240 & 2.91 & 2.4 &  12 \\
   \bottomrule
\end{tabular}
}
\end{center}
\end{table}
        The list of top players in $OPM/60$ is given in Table \ref{opm60p}.  The players in this list are are regarded by many as being among the best offensive players in the game, with the exception of a few players with low minutes played and higher errors: Alexander Radulov, Jakub Voracek, and Carl Gunnarsson.  Those players have far fewer minutes than the other players, and their estimates are less reliable.  Interestingly, Henrik Sedin actually has the second highest standard error in the list, even higher than Radulov and Voracek, despite the fact that he has much higher minutes played totals.  This is likely due to the fact that he spends most of his time playing with his brother Daniel (see Section \ref{errors}).
    \begin{figure}[h!] \centering
        \caption[Kernel Density Estimation for $OPM/60$ and $OPM$]
                {Kernel Density Estimation for $OPM/60$ Estimates and $OPM$ Estimates.}\label{opmfig}
        \includegraphics[width=.9\textwidth]{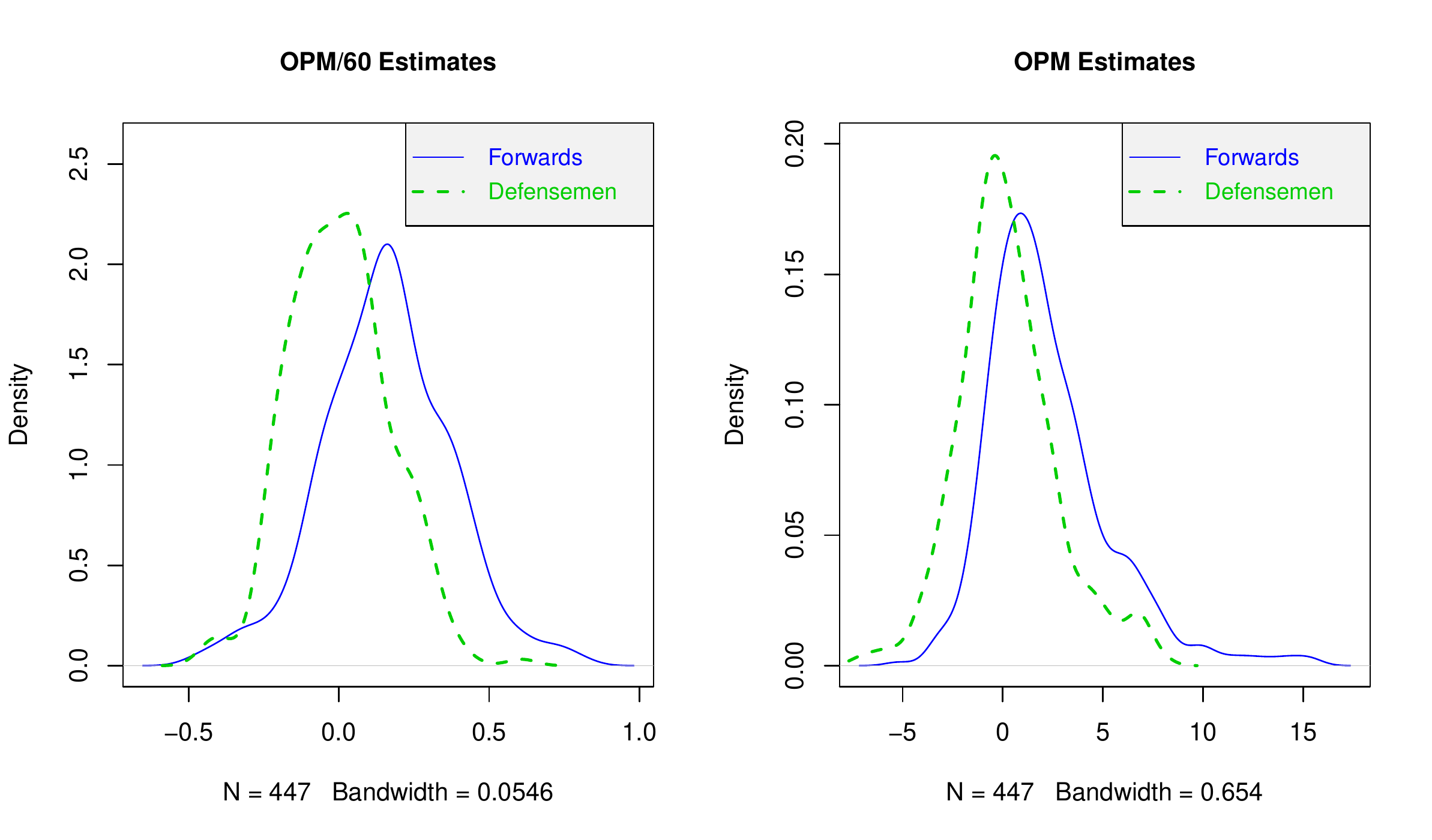}
    \end{figure}

        Forwards dominated the list in Table \ref{opm60p}.  That forwards are more prevalent than defensemen on this list is not unexpected, as one would probably assume that forwards contribute to offense more than defensemen do.  We can see this trend more clearly by plotting the kernel density estimation for $OPM/60$ for both forwards and defensemen.  This plot gives us an approximation of the histogram of our $OPM/60$ estimates for forwards and defensemen.  See Figure \ref{opmfig}.  The curve for forwards lies to the right of the curve for defensemen, suggesting that the $OPM/60$ estimates for forwards are generally higher than the $OPM/60$ estimates for defensemen.

        Since forwards dominated the list of top players in $OPM/60$, we give the top defensemen in $OPM/60$ in Table \ref{opm60d}.
\begin{table}[h!]
\begin{center}
\caption{
Top 10 Defensemen in OPM60 (minimum 700 minutes)
}
\label{opm60d}
{\small
\begin{tabular}{llrrrrrrrrr}
  \addlinespace[.3em] \toprule
Rk & Player & Pos & OPM60 & OErr & DPM60 & APM60 & Mins & GF60 & OPM & GF \\
  \midrule
 10 & C. Gunnarsson & D & 0.608 & 0.245 & 0.233 & 0.841 & 240 & 2.91 & 2.4 &  12 \\
   46 & Ville Koistinen & D & 0.424 & 0.210 & 0.082 & 0.506 & 380 & 2.89 & 2.7 &  18 \\
   71 & Andrei Markov & D & 0.370 & 0.162 & 0.036 & 0.405 & 1114 & 2.69 & 6.9 &  50 \\
   79 & Mike Green & D & 0.357 & 0.146 & 0.195 & 0.552 & 1334 & 3.30 & 7.9 &  73 \\
   82 & Mark Streit & D & 0.353 & 0.136 & $-$0.060 & 0.293 & 1199 & 2.60 & 7.1 &  52 \\
   95 & Johnny Oduya & D & 0.342 & 0.143 & 0.116 & 0.458 & 1209 & 2.78 & 6.9 &  56 \\
  112 & S. Robidas & D & 0.313 & 0.147 & 0.115 & 0.428 & 1316 & 2.60 & 6.9 &  57 \\
  113 & Ian White & D & 0.311 & 0.127 & 0.028 & 0.338 & 1343 & 2.73 & 6.9 &  61 \\
  119 & S. Brookbank & D & 0.302 & 0.164 & 0.163 & 0.465 & 640 & 2.44 & 3.2 &  26 \\
  122 & Bret Hedican & D & 0.297 & 0.166 & $-$0.162 & 0.135 & 594 & 2.96 & 2.9 &  29 \\
   \bottomrule
\end{tabular}
}
\end{center}
\end{table}
        There are some top offensive defensemen in this list, along with some players with low minutes, high errors, and skeptical ratings.  One example is Carl Gunnarsson, who tops this list and is one of the players with low minutes.  Gunnarsson did have a decent season offensively in 2009-10, scoring 12 even-strength points in 43 games, while playing just the 6th most minutes per game among defensemen on his team.  Projecting his statistics over 82 games, Gunnarsson would have had 23 even-strength points, tying him with Tomas Kaberle for the team lead among defensemen, despite playing less minutes.  So we do get an idea of why the model gave him this high estimate.  We note that the lower end of the 95\% confidence interval for Gunnarsson's $OPM/60$ is 0.118, suggesting that, at worst, he was still an above average offensive defenseman at even-strength during the limited minutes that he played.

 \subsection{$OPM$} Recall that $OPM$ is a measure of the offensive contribution of a player at even-strength in terms of goals over an entire season.  Once again, we list only forwards and defensemen in this section.
    The top 10 players in $OPM$ were already given and discussed in the introduction, Table \ref{opmp}.  That list was dominated by forwards, a trend that can also be seen in Figure \ref{opmfig}, so we now discuss the top 10 defensemen in $OPM$ given in Table \ref{opmd}.
\begin{table}[h!]
\begin{center}
\caption{
Top 10 Defensemen in OPM
}
\label{opmd}
{\small
\begin{tabular}{llrrrrrrrrr}
  \addlinespace[.3em] \toprule
Rk & Player & Pos & OPM & OErr & DPM & APM & Mins & GF60 & OPM60 & GF \\
  \midrule
 22 & Mike Green & D & 7.9 & 3.2 & 4.3 & 12.3 & 1334 & 3.30 & 0.357 &  73 \\
   31 & Mark Streit & D & 7.1 & 2.7 & $-$1.2 & 5.9 & 1199 & 2.60 & 0.353 &  52 \\
   35 & Andrei Markov & D & 6.9 & 3.0 & 0.7 & 7.5 & 1114 & 2.69 & 0.370 &  50 \\
   37 & Ian White & D & 6.9 & 2.8 & 0.6 & 7.6 & 1343 & 2.73 & 0.311 &  61 \\
   39 & S. Robidas & D & 6.9 & 3.2 & 2.5 & 9.4 & 1316 & 2.60 & 0.313 &  57 \\
   40 & Johnny Oduya & D & 6.9 & 2.9 & 2.3 & 9.2 & 1209 & 2.78 & 0.342 &  56 \\
   44 & Zdeno Chara & D & 6.6 & 3.9 & 2.2 & 8.8 & 1441 & 2.64 & 0.276 &  63 \\
   48 & Dion Phaneuf & D & 6.4 & 3.2 & $-$0.7 & 5.6 & 1443 & 2.76 & 0.265 &  66 \\
   53 & Duncan Keith & D & 6.2 & 4.2 & 7.3 & 13.5 & 1532 & 2.94 & 0.245 &  75 \\
   67 & Dan Boyle & D & 5.6 & 2.7 & $-$1.6 & 4.0 & 1169 & 2.69 & 0.286 &  52 \\
   \bottomrule
\end{tabular}
}
\end{center}
\end{table}
    Most of the players in Table \ref{opmd} are among the top offensive defensemen in the league at even-strength.  Nicklas Lidstrom is one notable omission.  Lidstrom is 11th among defensemen with an $OPM$ of 5.5.  Interestingly, the Ilardi-Barzilai model estimates a 3.8 $OPM$ for Lidstrom, while the Rosenbaum-type model, with goalies included on offensive, estimates a 7.3 $OPM$.  It seems that including, or not including, goalies on offense has a big effect on Lidstrom's estimate.  It turns out that other Detroit Red Wings skaters are affected also.  We discuss goalies and offense, and the effect it had on the Detroit Red Wings, as well as the New York Rangers, in Section \ref{goalies}.

        \subsection{$DPM/60$}\label{dpm60}   
\begin{figure}[h!] \centering
                \caption[Kernel Density Estimation for $DPM/60$ and $DPM$]
                {Kernel Density Estimation for $DPM/60$ Estimates and $DPM$ Estimates.}\label{dpmfig}
                \includegraphics[width=.9\textwidth]{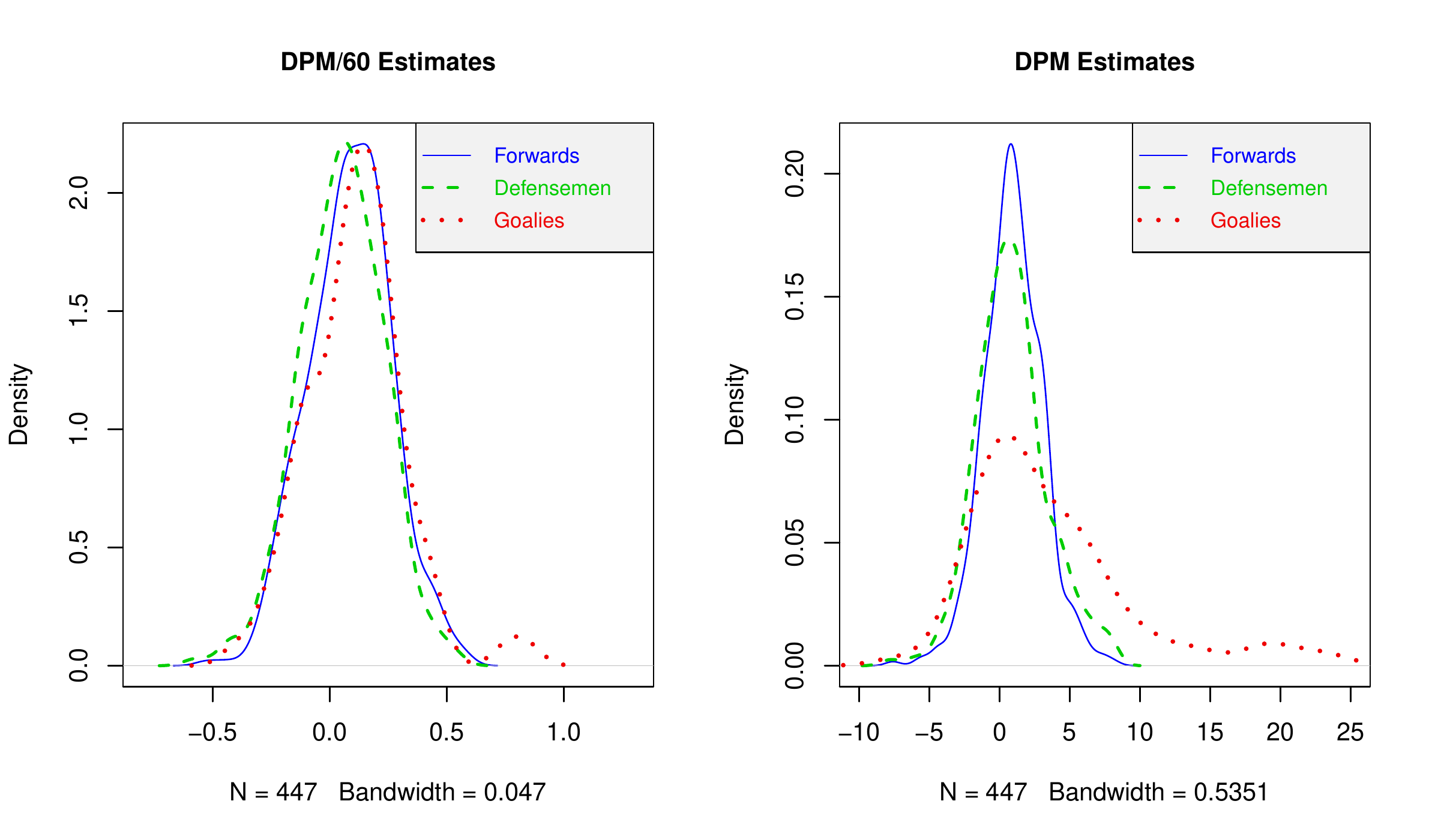}
            \end{figure}

Recall that $DPM/60$ is a measure of the defensive contribution of a player in terms of goals per 60 minutes of playing time at even-strength.
\begin{table}[h!]
\begin{center}
\caption{
Top 10 Players in DPM60
}
\label{dpm60-no-mins}
{\small
\begin{tabular}{llrrrrrrrrr}
  \addlinespace[.3em] \toprule
Rk & Player & Pos & OPM60 & DPM60 & DErr & APM60 & Mins & GA60 & DPM & GA \\
  \midrule
763 & Pekka Rinne & G & NA & 0.845 & 0.232 & 0.845 & 1680 & 2.12 & 23.7 &  59 \\
  717 & Dan Ellis & G & NA & 0.757 & 0.218 & 0.757 & 1509 & 2.32 & 19.0 &  58 \\
  433 & George Parros & RW & 0.035 & 0.576 & 0.220 & 0.611 & 387 & 0.98 & 3.7 &   6 \\
  424 & Derek Dorsett & RW & 0.040 & 0.571 & 0.225 & 0.611 & 317 & 1.45 & 3.0 &   8 \\
  166 & Peter Regin & C & 0.242 & 0.531 & 0.229 & 0.773 & 324 & 1.73 & 2.9 &   9 \\
  688 & Adam Hall & RW & $-$0.336 & 0.528 & 0.211 & 0.192 & 329 & 1.58 & 2.9 &   9 \\
  505 & Paul Martin & D & $-$0.026 & 0.526 & 0.170 & 0.500 & 916 & 1.55 & 8.0 &  24 \\
  336 & Mark Fistric & D & 0.107 & 0.510 & 0.173 & 0.617 & 594 & 1.48 & 5.1 &  15 \\
  456 & Drew Miller & LW & 0.021 & 0.490 & 0.205 & 0.510 & 370 & 1.62 & 3.0 &  10 \\
  156 & Josef Vasicek & C & 0.251 & 0.484 & 0.253 & 0.735 & 343 & 1.69 & 2.8 &  10 \\
   \bottomrule
\end{tabular}
}
\end{center}
\end{table}
 Without specifying a minimum minutes played limit, we get two goalies, then several players with low minutes, in the list of top players in $DPM/60$ given in Table \ref{dpm60-no-mins}.  In order to remove the players with low minutes from this list, we restrict the list to those players with more than 700 minutes played.  The new list is given in Table \ref{dpm60p}.
\begin{table}[h!]
\begin{center}
\caption{
Top 10 Players in DPM60 (minimum 700 minutes)
}
\label{dpm60p}
{\small
\begin{tabular}{llrrrrrrrrr}
  \addlinespace[.3em] \toprule
Rk & Player & Pos & OPM60 & DPM60 & DErr & APM60 & Mins & GA60 & DPM & GA \\
  \midrule
  1 & Pekka Rinne & G & NA & 0.845 & 0.232 & 0.845 & 1680 & 2.12 & 23.7 &  59 \\
    2 & Dan Ellis & G & NA & 0.757 & 0.218 & 0.757 & 1509 & 2.32 & 19.0 &  58 \\
    7 & Paul Martin & D & $-$0.026 & 0.526 & 0.170 & 0.500 & 916 & 1.55 & 8.0 &  24 \\
   12 & Mikko Koivu & C & 0.383 & 0.469 & 0.176 & 0.852 & 1032 & 2.02 & 8.1 &  35 \\
   14 & Chris Mason & G & NA & 0.460 & 0.169 & 0.460 & 2384 & 2.31 & 18.3 &  92 \\
   15 & Ryan Callahan & RW & 0.022 & 0.453 & 0.153 & 0.475 & 878 & 1.78 & 6.6 &  26 \\
   18 & Marco Sturm & LW & 0.169 & 0.439 & 0.179 & 0.608 & 702 & 1.57 & 5.1 &  18 \\
   22 & Jason Pominville & RW & 0.309 & 0.424 & 0.176 & 0.733 & 1052 & 2.30 & 7.4 &  40 \\
   24 & Marty Turco & G & NA & 0.424 & 0.154 & 0.424 & 2787 & 2.27 & 19.7 & 105 \\
   27 & Tomas Plekanec & C & 0.004 & 0.411 & 0.168 & 0.414 & 1053 & 2.13 & 7.2 &  37 \\
   \bottomrule
\end{tabular}
}
\end{center}
\end{table}
 In that list we get a mix of goalies, forwards and defensemen.   To see if this trend continues outside the top 10, we can again plot a kernel density estimation of $DPM/60$ estimates for forwards, defensemen, and goalies.  See Figure \ref{dpmfig}.
Forwards, defensemen, and goalies seem to have a fairly similar distribution of $DPM/60$ estimates, though defensemen may be slightly behind forwards and goalies.  It may seem counterintuitive that defensemen have lower ratings than forwards.  The estimates seem to indicate that forwards contribute more to defense (per 60 minutes of ice time) than defensemen do, but the difference in estimates is so small that it could be simply due to noise.  The trends are slightly different with $DPM$, which is playing-time dependent.  The top goalies in $DPM/60$ typically play more minutes than forwards and defensemen, and their $DPM$'s are much higher.  Top defensemen typically play more minutes than top forwards, so their ratings are helped when playing-time is considered as well.

We now discuss the top 10 in $DPM/60$ for skaters, forwards, defensemen, and goalies, separately.
\begin{table}[h!]
\begin{center}
\caption{
Top 10 Skaters in DPM60 (minimum 700 minutes)
}
\label{dpm60s}
{\small
\begin{tabular}{llrrrrrrrrr}
  \addlinespace[.3em] \toprule
Rk & Player & Pos & OPM60 & DPM60 & DErr & APM60 & Mins & GA60 & DPM & GA \\
  \midrule
  7 & Paul Martin & D & $-$0.026 & 0.526 & 0.170 & 0.500 & 916 & 1.55 & 8.0 &  24 \\
   12 & Mikko Koivu & C & 0.383 & 0.469 & 0.176 & 0.852 & 1032 & 2.02 & 8.1 &  35 \\
   15 & Ryan Callahan & RW & 0.022 & 0.453 & 0.153 & 0.475 & 878 & 1.78 & 6.6 &  26 \\
   18 & Marco Sturm & LW & 0.169 & 0.439 & 0.179 & 0.608 & 702 & 1.57 & 5.1 &  18 \\
   22 & Jason Pominville & RW & 0.309 & 0.424 & 0.176 & 0.733 & 1052 & 2.30 & 7.4 &  40 \\
   27 & Tomas Plekanec & C & 0.004 & 0.411 & 0.168 & 0.414 & 1053 & 2.13 & 7.2 &  37 \\
   31 & Willie Mitchell & D & $-$0.122 & 0.404 & 0.143 & 0.281 & 1178 & 1.90 & 7.9 &  37 \\
   37 & Manny Malhotra & C & 0.167 & 0.380 & 0.142 & 0.547 & 922 & 1.82 & 5.8 &  28 \\
   38 & Andrew Greene & D & $-$0.118 & 0.379 & 0.161 & 0.262 & 1001 & 1.70 & 6.3 &  28 \\
   39 & Jan Hejda & D & 0.061 & 0.373 & 0.158 & 0.434 & 1269 & 2.24 & 7.9 &  47 \\
   \bottomrule
\end{tabular}
}
\end{center}
\end{table}
A mix of defensive forwards and defensive defensemen appear in the list of top skaters in Table \ref{dpm60s}.  Paul Martin leads this list by a sizeable margin, partly because of his very low $GA/60$ of 1.55.  Martin is followed by 5 forwards, including Marco Sturm.  Sturm has just above the minimum for minutes played, but since he has a $GA/60$ of 1.57 during his limited playing time, his estimate is believable.

\begin{table}[h!]
\begin{center}
\caption{
Top 10 Forwards in DPM60 (minimum 700 minutes)
}
\label{dpm60f}
{\small
\begin{tabular}{llrrrrrrrrr}
  \addlinespace[.3em] \toprule
Rk & Player & Pos & OPM60 & DPM60 & DErr & APM60 & Mins & GA60 & DPM & GA \\
  \midrule
 12 & Mikko Koivu & C & 0.383 & 0.469 & 0.176 & 0.852 & 1032 & 2.02 & 8.1 &  35 \\
   15 & Ryan Callahan & RW & 0.022 & 0.453 & 0.153 & 0.475 & 878 & 1.78 & 6.6 &  26 \\
   18 & Marco Sturm & LW & 0.169 & 0.439 & 0.179 & 0.608 & 702 & 1.57 & 5.1 &  18 \\
   22 & Jason Pominville & RW & 0.309 & 0.424 & 0.176 & 0.733 & 1052 & 2.30 & 7.4 &  40 \\
   27 & Tomas Plekanec & C & 0.004 & 0.411 & 0.168 & 0.414 & 1053 & 2.13 & 7.2 &  37 \\
   37 & Manny Malhotra & C & 0.167 & 0.380 & 0.142 & 0.547 & 922 & 1.82 & 5.8 &  28 \\
   41 & Tyler Kennedy & C & 0.343 & 0.365 & 0.163 & 0.708 & 730 & 1.75 & 4.4 &  21 \\
   42 & David Krejci & C & 0.281 & 0.361 & 0.189 & 0.642 & 912 & 1.89 & 5.5 &  29 \\
   44 & Travis Moen & LW & $-$0.323 & 0.350 & 0.152 & 0.028 & 969 & 1.84 & 5.7 &  30 \\
   45 & Daniel Sedin & LW & 0.364 & 0.346 & 0.231 & 0.710 & 1057 & 2.03 & 6.1 &  36 \\
   \bottomrule
\end{tabular}
}
\end{center}
\end{table}
In the list of top forwards in $DPM/60$ given in Table \ref{dpm60f}, we get some well-known defensive forwards, and a couple interesting names.  Once again a Sedin twin has the highest errors in a list.  The model seems to be giving Daniel the credit for the Sedin line's success in $DPM/60$, whereas for $OPM/60$, Henrik gets the credit.  Henrik's $DPM/60$ estimate is actually negative ($-0.294$), while his $OPM/60$ estimate (0.718) is much higher than Daniel's $OPM/60$ estimate (0.364).

\begin{table}[b!]
\begin{center}
\caption{
Top 10 Defensemen in DPM60 (minimum 700 minutes)
}
\label{dpm60d}
{\small
\begin{tabular}{llrrrrrrrrr}
  \addlinespace[.3em] \toprule
Rk & Player & Pos & OPM60 & DPM60 & DErr & APM60 & Mins & GA60 & DPM & GA \\
  \midrule
  7 & Paul Martin & D & $-$0.026 & 0.526 & 0.170 & 0.500 & 916 & 1.55 & 8.0 &  24 \\
   31 & Willie Mitchell & D & $-$0.122 & 0.404 & 0.143 & 0.281 & 1178 & 1.90 & 7.9 &  37 \\
   38 & Andrew Greene & D & $-$0.118 & 0.379 & 0.161 & 0.262 & 1001 & 1.70 & 6.3 &  28 \\
   39 & Jan Hejda & D & 0.061 & 0.373 & 0.158 & 0.434 & 1269 & 2.24 & 7.9 &  47 \\
   46 & Mike Weaver & D & $-$0.419 & 0.341 & 0.153 & $-$0.078 & 819 & 1.73 & 4.7 &  24 \\
   62 & Nicklas Lidstrom & D & 0.242 & 0.307 & 0.191 & 0.549 & 1374 & 1.82 & 7.0 &  42 \\
   67 & Tobias Enstrom & D & $-$0.005 & 0.301 & 0.166 & 0.296 & 1319 & 2.70 & 6.6 &  59 \\
   68 & Sean O'donnell & D & 0.027 & 0.298 & 0.133 & 0.325 & 1180 & 1.75 & 5.9 &  34 \\
   69 & Mike Lundin & D & $-$0.035 & 0.297 & 0.157 & 0.263 & 738 & 2.14 & 3.7 &  26 \\
   70 & Marc-E Vlasic & D & 0.030 & 0.296 & 0.150 & 0.326 & 1311 & 1.89 & 6.5 &  41 \\
   \bottomrule
\end{tabular}
}
\end{center}
\end{table}
Tyler Kennedy is part of what many hockey analysts consider to be one of the top defensive lines in hockey, and his $DPM/60$ estimate supports that belief.  We note that Jordan Staal's $DPM/60$ estimate is $.207 \pm .151$, and he has a $GA/60$ of 1.95, so we can see one possible reason why the model gave Kennedy the higher estimate of the two linemates.  Incidentally, if we weight the observations in the model so that the 2009-2010 season counts more heavily than the other two seasons, Staal makes the list of top 10 forwards in $DPM/60$.  This estimate supports his nomination for the 2009-2010 Selke trophy, which is given each season to the top defensive forward in the game.

Mike Weaver and Mike Lundin are members of the list of top defensemen in $DPM/60$, which is shown in Table \ref{dpm60d}.  Weaver has the second lowest $GA/60$ in this list at 1.73, and his most common teammates, Chris Mason (2.31) and Carlo Colaiacovo (2.41) have a higher $GA/60,$ so we see one reason why the model gave him a high $DPM/60$ rating.  Similarly, Lundin's $GA/60$, while not as low as Weaver's, is still fairly low, despite the fact that his most common teammates have a very high $GA/60$ (Mike Smith, 2.44; Vincent Lecavalier, 3.03; Martin St. Louis, 2.97).  Also, according to Gabriel Desjardins' Quality of Competition (QualComp) statistic from \cite{gabe}, Lundin had the highest QualComp in 2009-2010 among players with at least 10 games played, indicating that he performed well against strong competition.
\begin{table}[h!]
\begin{center}
\caption{
Top 10 Goalies in DPM60 (minimum 700 minutes)
}
\label{dpm60g}
{\small
\begin{tabular}{llrrrrrrrrr}
  \addlinespace[.3em] \toprule
Rk & Player & Pos & OPM60 & DPM60 & DErr & APM60 & Mins & GA60 & DPM & GA \\
  \midrule
  1 & Pekka Rinne & G & NA & 0.845 & 0.232 & 0.845 & 1680 & 2.12 & 23.7 &  59 \\
    2 & Dan Ellis & G & NA & 0.757 & 0.218 & 0.757 & 1509 & 2.32 & 19.0 &  58 \\
   14 & Chris Mason & G & NA & 0.460 & 0.169 & 0.460 & 2384 & 2.31 & 18.3 &  92 \\
   24 & Marty Turco & G & NA & 0.424 & 0.154 & 0.424 & 2787 & 2.27 & 19.7 & 105 \\
   29 & Erik Ersberg & G & NA & 0.410 & 0.199 & 0.410 & 728 & 2.14 & 5.0 &  26 \\
   30 & H. Lundqvist & G & NA & 0.410 & 0.220 & 0.410 & 3223 & 2.10 & 22.0 & 113 \\
   34 & Jonathan Quick & G & NA & 0.394 & 0.187 & 0.394 & 1781 & 2.15 & 11.7 &  64 \\
   54 & Cam Ward & G & NA & 0.317 & 0.151 & 0.317 & 2655 & 2.23 & 14.0 &  99 \\
   75 & Tuukka Rask & G & NA & 0.293 & 0.259 & 0.293 & 763 & 1.70 & 3.7 &  22 \\
   76 & Ty Conklin & G & NA & 0.293 & 0.178 & 0.293 & 1392 & 2.13 & 6.8 &  49 \\
   \bottomrule
\end{tabular}
}
\end{center}
\end{table}

 Interestingly, the top 3 goalies in $DPM/60$, as shown in Table \ref{dpm60g}, have played for the Nashville Predators at some point during the past three seasons.   There are a couple possible reasons for this trend.
One reason is that the estimates are noisy, so it could simply be a coincidence that those three goalies ended up at the top of this list.  The estimates for Rinne and Ellis are significantly higher than those of the other goalies, but several other goalies are within one standard error of the top 3 in $DPM/60$, and a few are within two standard errors of the top spot.  Note that the low end of the 95\% confidence intervals of the $DPM/60$ estimates for Rinne and Ellis are still in the top 7 in $DPM/60$, suggesting that, at worst, they were still very good.

Even if the goalies' $DPM/60$ estimates were not noisy, $DPM/60$ would still not be the best way to isolate and measure goalie's individual ability.  Recall that the interpretation of a goalie's $DPM/60$ is goals per 60 minutes contributed by the goalie on defense, or goals per 60 minutes prevented by the goalie.  We could think of $DPM/60$ as measuring the difference between a goalie's goals against average at even strength and the league's goals against average at even strength, while adjusting for the strength of the teammates and opponents of the goalie.  A goalie who has a relatively low goals against average at even strength should in general have a relatively high (good) $DPM/60$ estimate.  In general, this relationship is true for our results.  In Table \ref{dpm60g}, the 10 goalies with the highest $DPM/60$ estimates also have low $GA/60$ statistics.

Unfortunately, goals against average is not the best measure of a goalie's ability.  The number of goals per 60 minutes allowed by a team depends on not only the goalie's ability at stopping shots on goal, but also the frequency and quality of the shots on goal that his team allows.  So goals against average is a measure not just of a goalie's ability, but also of his team's ability at preventing shots on goal.  Ideally, the model would be able to correctly determine if the goalie or the team in front of him deserves credit for a low goals against average, but that does not seem to be happening.  One reason could be the relatively low number of goalies on each team.  Another reason could be that there is some team-level effect not accounted for.  If we include team variables in the model, the results are even worse.  The team estimates are very noisy (with errors around 0.50), the goalie estimates are even noisier with the team variables than without the team variables, and the model still does not isolate a goalie's ability.

Different techniques for measuring a goalie's ability and contribution to his team would be preferred over $DPM/60$.  Most methods would likely use different information, including the quality and frequency of the shots on goal that his team allows.  See, for example, Ken Krzywicki's shot quality model in \cite{ken1} and \cite{ken2}.

\subsection{$DPM$}\label{dpm}
Recall that $DPM$ is a measure of the defensive contribution of a player at even-strength in terms of goals over an entire season.  We now discuss the top 10 players, skaters, forwards, and defensemen in $DPM$.
\begin{table}[h!]
\begin{center}
\caption{
Top 10 Players in DPM
}
\label{dpmp}
{\small
\begin{tabular}{llrrrrrrrrr}
  \addlinespace[.3em] \toprule
Rk & Player & Pos & OPM & DPM & DErr & APM & Mins & GA60 & DPM60 & GA \\
  \midrule
  1 & Pekka Rinne & G & NA & 23.7 & 6.5 & 23.7 & 1680 & 2.12 & 0.845 &  59 \\
    2 & H. Lundqvist & G & NA & 22.0 & 11.8 & 22.0 & 3223 & 2.10 & 0.410 & 113 \\
    3 & Marty Turco & G & NA & 19.7 & 7.1 & 19.7 & 2787 & 2.27 & 0.424 & 105 \\
    4 & Dan Ellis & G & NA & 19.0 & 5.5 & 19.0 & 1509 & 2.32 & 0.757 &  58 \\
    5 & Chris Mason & G & NA & 18.3 & 6.7 & 18.3 & 2384 & 2.31 & 0.460 &  92 \\
    6 & Ryan Miller & G & NA & 14.2 & 9.6 & 14.2 & 3078 & 2.29 & 0.276 & 118 \\
    7 & Cam Ward & G & NA & 14.0 & 6.7 & 14.0 & 2655 & 2.23 & 0.317 &  99 \\
    8 & Jonathan Quick & G & NA & 11.7 & 5.6 & 11.7 & 1781 & 2.15 & 0.394 &  64 \\
    9 & Tomas Vokoun & G & NA & 10.2 & 8.8 & 10.2 & 2885 & 2.22 & 0.211 & 107 \\
   10 & Ilja Bryzgalov & G & NA & 10.0 & 7.6 & 10.0 & 2942 & 2.17 & 0.203 & 106 \\
   \bottomrule
\end{tabular}
}
\end{center}
\end{table}
 The list of top players in $DPM$ given in Table \ref{dpmp} is entirely made up of goalies, which is not unexpected.  Many people consider goalie the most important position in hockey, and this list seems to support that claim, at least for the defensive component of the game.  While many goalies have a lower $DPM/60$ than many skaters, the comparatively high minutes played for goalies bump many of them to the top of the list in $DPM$.  Another consequence of the high minutes played is that the standard errors for goalies are now very high for $DPM$.  This fact makes the $DPM$ estimates for goalies less reliable than the $DPM$ estimates for skaters.  We reiterate what we discussed at the end of Section \ref{dpm60}: other methods of rating goalies are preferred over $DPM/60$ and $DPM$.
\begin{table}[h!]
\begin{center}
\caption{
Top 10 Skaters in DPM
}
\label{dpms}
{\small
\begin{tabular}{llrrrrrrrrr}
  \addlinespace[.3em] \toprule
Rk & Player & Pos & OPM & DPM & DErr & APM & Mins & GA60 & DPM60 & GA \\
  \midrule
 13 & Mikko Koivu & C & 6.6 & 8.1 & 3.0 & 14.7 & 1032 & 2.02 & 0.469 &  35 \\
   14 & Paul Martin & D & $-$0.4 & 8.0 & 2.6 & 7.6 & 916 & 1.55 & 0.526 &  24 \\
   15 & Jan Hejda & D & 1.3 & 7.9 & 3.3 & 9.2 & 1269 & 2.24 & 0.373 &  47 \\
   16 & Willie Mitchell & D & $-$2.4 & 7.9 & 2.8 & 5.5 & 1178 & 1.90 & 0.404 &  37 \\
   18 & Jay Bouwmeester & D & $-$3.1 & 7.5 & 3.2 & 4.4 & 1532 & 2.17 & 0.292 &  55 \\
   19 & Jason Pominville & RW & 5.4 & 7.4 & 3.1 & 12.9 & 1052 & 2.30 & 0.424 &  40 \\
   20 & Duncan Keith & D & 6.2 & 7.3 & 4.2 & 13.5 & 1532 & 2.21 & 0.284 &  56 \\
   21 & Tomas Plekanec & C & 0.1 & 7.2 & 3.0 & 7.3 & 1053 & 2.13 & 0.411 &  37 \\
   23 & Nicklas Lidstrom & D & 5.5 & 7.0 & 4.4 & 12.6 & 1374 & 1.82 & 0.307 &  42 \\
   26 & Ryan Callahan & RW & 0.3 & 6.6 & 2.2 & 6.9 & 878 & 1.78 & 0.453 &  26 \\
   \bottomrule
\end{tabular}
}
\end{center}
\end{table}

Unlike the list of top skaters in $DPM/60$ (Table \ref{dpm60s}), defensemen are more prevalent than forwards on the list of top skaters in $DPM$ given in Table \ref{dpms}, due to their higher minutes played.  Defensemen make up 5 of the first 7, and 9 of the first 13 skaters in $DPM$.  Beyond the top 13, the distribution of $DPM$ estimates for forwards are actually very similar (see Figure \ref{dpmfig}).
\begin{table}[h!]
\begin{center}
\caption{
Top 10 Forwards in DPM
}
\label{dpmf}
{\small
\begin{tabular}{llrrrrrrrrr}
  \addlinespace[.3em] \toprule
Rk & Player & Pos & OPM & DPM & DErr & APM & Mins & GA60 & DPM60 & GA \\
  \midrule
 13 & Mikko Koivu & C & 6.6 & 8.1 & 3.0 & 14.7 & 1032 & 2.02 & 0.469 &  35 \\
   19 & Jason Pominville & RW & 5.4 & 7.4 & 3.1 & 12.9 & 1052 & 2.30 & 0.424 &  40 \\
   21 & Tomas Plekanec & C & 0.1 & 7.2 & 3.0 & 7.3 & 1053 & 2.13 & 0.411 &  37 \\
   26 & Ryan Callahan & RW & 0.3 & 6.6 & 2.2 & 6.9 & 878 & 1.78 & 0.453 &  26 \\
   33 & Pavel Datsyuk & C & 15.4 & 6.2 & 3.5 & 21.6 & 1186 & 1.84 & 0.314 &  36 \\
   35 & Daniel Sedin & LW & 6.4 & 6.1 & 4.1 & 12.5 & 1057 & 2.03 & 0.346 &  36 \\
   39 & Manny Malhotra & C & 2.6 & 5.8 & 2.2 & 8.4 & 922 & 1.82 & 0.380 &  28 \\
   41 & D. Langkow & C & $-$0.4 & 5.7 & 2.8 & 5.2 & 1010 & 2.02 & 0.336 &  34 \\
   43 & Travis Moen & LW & $-$5.2 & 5.7 & 2.5 & 0.4 & 969 & 1.84 & 0.350 &  30 \\
   45 & David Krejci & C & 4.3 & 5.5 & 2.9 & 9.8 & 912 & 1.89 & 0.361 &  29 \\
   \bottomrule
\end{tabular}
}
\end{center}
\end{table}

We now look at forwards and defensemen separately.  Many of the top forwards in Table \ref{dpmf} are known to be very solid defensive forwards.  Mikko Koivu is often praised for his work defensively, and Pavel Datsyuk is a two-time Selke Trophy winner for the best defensive forward in the league.  Jason Pominville's ranking is surprising given that his $GA/60$ is the worst among players on this list.  Checking his most common linemates, we find Ryan Miller (2.29 $GA/60$), Jochen Hecht (2.64), and Toni Lydman (2.36), whose $GA/60$ are not significantly different than Pominville's.  Pominville did lead his team in traditional plus-minus in 2007-2008 (+16), and was tied for second in 2009-2010 (+13), which may have caused the high rating, but he was also a $-4$ in 2008-2009.
\begin{table}[h!]
\begin{center}
\caption{
Top 10 Defensemen in DPM
}
\label{dpmd}
{\small
\begin{tabular}{llrrrrrrrrr}
  \addlinespace[.3em] \toprule
Rk & Player & Pos & OPM & DPM & DErr & APM & Mins & GA60 & DPM60 & GA \\
  \midrule
 14 & Paul Martin & D & $-$0.4 & 8.0 & 2.6 & 7.6 & 916 & 1.55 & 0.526 &  24 \\
   15 & Jan Hejda & D & 1.3 & 7.9 & 3.3 & 9.2 & 1269 & 2.24 & 0.373 &  47 \\
   16 & Willie Mitchell & D & $-$2.4 & 7.9 & 2.8 & 5.5 & 1178 & 1.90 & 0.404 &  37 \\
   18 & Jay Bouwmeester & D & $-$3.1 & 7.5 & 3.2 & 4.4 & 1532 & 2.17 & 0.292 &  55 \\
   20 & Duncan Keith & D & 6.2 & 7.3 & 4.2 & 13.5 & 1532 & 2.21 & 0.284 &  56 \\
   23 & Nicklas Lidstrom & D & 5.5 & 7.0 & 4.4 & 12.6 & 1374 & 1.82 & 0.307 &  42 \\
   27 & Tobias Enstrom & D & $-$0.1 & 6.6 & 3.7 & 6.5 & 1319 & 2.70 & 0.301 &  59 \\
   28 & Marc-E Vlasic & D & 0.7 & 6.5 & 3.3 & 7.1 & 1311 & 1.89 & 0.296 &  41 \\
   30 & Andrew Greene & D & $-$2.0 & 6.3 & 2.7 & 4.4 & 1001 & 1.70 & 0.379 &  28 \\
   36 & Ron Hainsey & D & $-$2.9 & 6.1 & 2.9 & 3.2 & 1264 & 2.50 & 0.289 &  53 \\
   \bottomrule
\end{tabular}
}
\end{center}
\end{table}

Paul Martin, the leader among skaters in $DPM/60$, tops the list of best defensemen in $DPM$ given in Table \ref{dpmd}, despite much lower minutes played than the others.  Martin has a nice list of most common teammates (Martin Brodeur, 1.92 $GA/60$; Johnny Oduya, 2.00; Zach Parise, 1.74) but his $GA/60$ is extremely low, which is probably the cause of his low $DPM/60$ and $DPM$ estimates.  Tobias Enstrom's $DPM/60$ and $DPM$ estimates are high given that his 2.70 $GA/60$ and 59 $GA$ statistics are the worst in the list.  Enstrom's $GA/60$ is not significantly different than the $GA/60$ of his most common linemates, Niclas Havelid (2.87 $GA/60$), Johan Hedberg (2.67), and Kari Lehtonen (2.63).  Further down Enstrom's list of common linemates is Ilya Kovalchuk, whose $3.09$ $GA/60$ and $-4.2$ $DPM$ are among the worst in the league.  All teammates (and opponents) affect the model's estimates, not just the three most common teammates, so Kovalchuk and some other teammates with low defensive abilities could be increasing Enstrom's defensive estimates.  Another Atlanta defensemen, Ron Hainsey, also has a high $DPM$ given his raw statistics.  Looking deeper, his 2.50 $GA/60$ is actually second best on his team among players with more than 700 minutes.  Our model seems to be saying that Hainsey, like Enstrom, is better than his raw statistics suggest, mostly because of the quality of teammates that he plays with.

    \subsection{$APM/60$}  We now begin to look at the top players in the league in terms of $APM/60$ and $APM$.  Recall that $APM/60$ is a measure of the total (offensive and defensive) contribution of a player at even-strength in terms of net goals (goals for minus goals against) per 60 minutes of playing time.
\begin{table}[h!]
\begin{center}
\caption{
Top 10 Players in APM60 (minimum 700 minutes)
}
\label{apm60p}
{\small
\begin{tabular}{llrrrrrrrrr}
  \addlinespace[.3em] \toprule
Rk & Player & Pos & OPM60 & DPM60 & APM60 & Err & Mins & NG60 & APM & NG \\
  \midrule
  1 & Pavel Datsyuk & C & 0.777 & 0.314 & 1.091 & 0.247 & 1186 & 1.55 & 21.6 &  31 \\
    2 & Marian Gaborik & RW & 0.715 & 0.303 & 1.018 & 0.222 & 853 & 1.01 & 14.5 &  15 \\
    3 & Mikko Koivu & C & 0.383 & 0.469 & 0.852 & 0.246 & 1032 & 0.52 & 14.7 &   9 \\
    4 & Pekka Rinne & G & NA & 0.845 & 0.845 & 0.232 & 1680 & NA & 23.7 & NA \\
    7 & Zach Parise & LW & 0.652 & 0.155 & 0.807 & 0.236 & 1164 & 1.20 & 15.6 &  23 \\
    9 & Joe Thornton & C & 0.590 & 0.177 & 0.767 & 0.227 & 1222 & 1.08 & 15.6 &  22 \\
   10 & Sidney Crosby & C & 0.818 & $-$0.052 & 0.766 & 0.212 & 1059 & 1.04 & 13.5 &  18 \\
   11 & Dan Ellis & G & NA & 0.757 & 0.757 & 0.218 & 1509 & NA & 19.0 & NA \\
   12 & Tim Connolly & C & 0.501 & 0.244 & 0.745 & 0.247 & 710 & 0.90 & 8.8 &  11 \\
   15 & Alex Ovechkin & LW & 0.723 & 0.010 & 0.733 & 0.254 & 1262 & 1.42 & 15.4 &  30 \\
   \bottomrule
\end{tabular}
}
\end{center}
\end{table}

\begin{figure}[h!] \centering
                \caption[Kernel Density Estimation for APM/60 and $APM$]
                {Kernel Density Estimation for APM/60 Estimates and $APM$ Estimates.}\label{apmfig}
                \includegraphics[width=.9\textwidth]{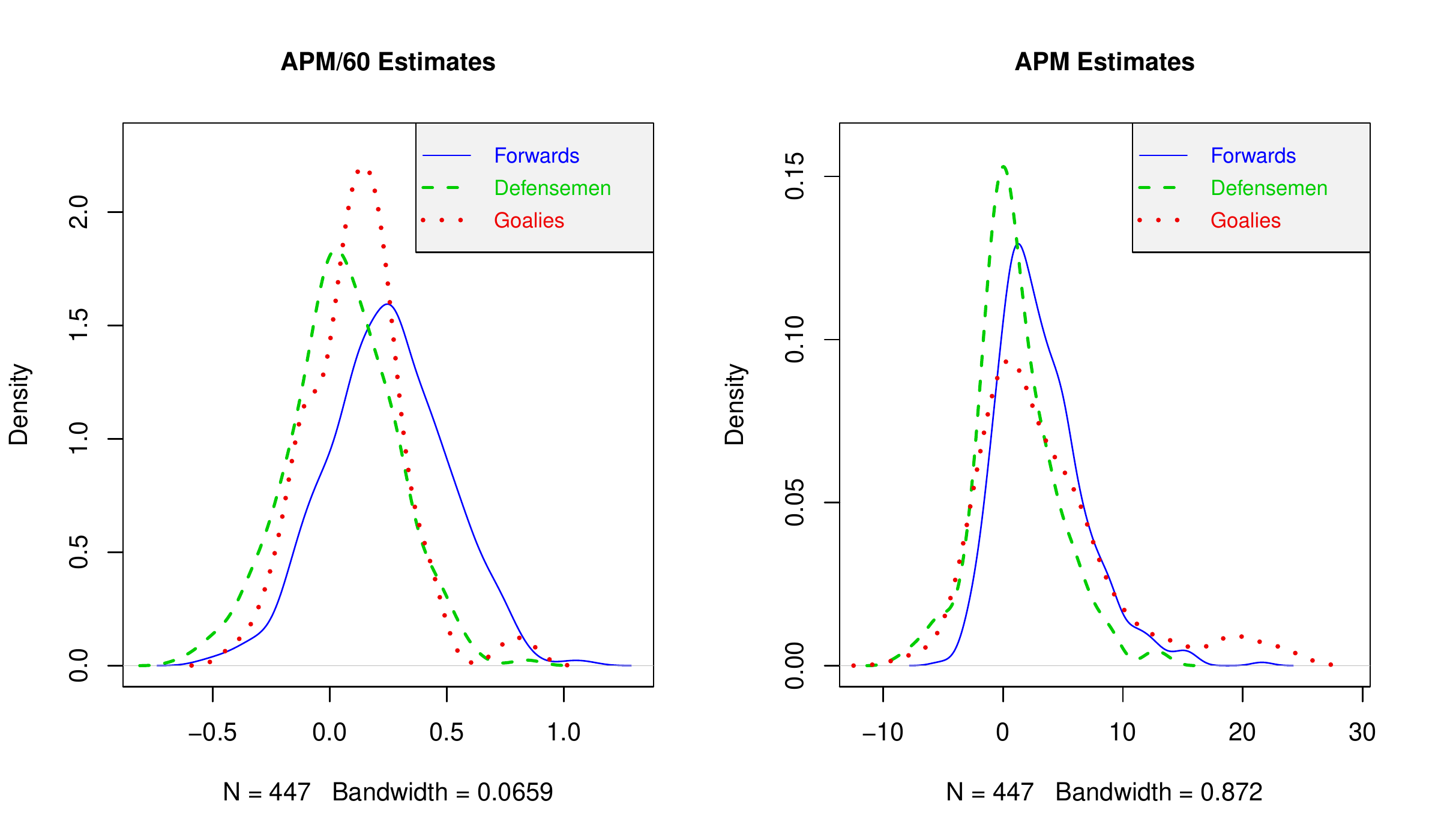}
            \end{figure}

Datysuk, considered by many to be the best two-way player in the game, tops the list of best players in $APM/60$ given in Table \ref{apm60p}.  Only two goalies make the top 10.  We plot the kernel density estimate for $APM/60$ and $APM$ in Figure \ref{apmfig} to get an idea of whether this trend continues outside the top 10.
Forwards seem to have higher estimates than goalies and defensemen.  Note that the picture changes slightly for $APM$, but defensemen still seem to have the lowest estimates in general.  Goalies seem to have the widest spread in $APM$, which is expected because of their high minutes played.

\begin{table}[h!]
\begin{center}
\caption{
Top 10 Forwards in APM60 (minimum 700 minutes)
}
\label{apm60f}
{\small
\begin{tabular}{llrrrrrrrrr}
  \addlinespace[.3em] \toprule
Rk & Player & Pos & OPM60 & DPM60 & APM60 & Err & Mins & NG60 & APM & NG \\
  \midrule
  1 & Pavel Datsyuk & C & 0.777 & 0.314 & 1.091 & 0.247 & 1186 & 1.55 & 21.6 &  31 \\
    2 & Marian Gaborik & RW & 0.715 & 0.303 & 1.018 & 0.222 & 853 & 1.01 & 14.5 &  15 \\
    3 & Mikko Koivu & C & 0.383 & 0.469 & 0.852 & 0.246 & 1032 & 0.52 & 14.7 &   9 \\
    7 & Zach Parise & LW & 0.652 & 0.155 & 0.807 & 0.236 & 1164 & 1.20 & 15.6 &  23 \\
    9 & Joe Thornton & C & 0.590 & 0.177 & 0.767 & 0.227 & 1222 & 1.08 & 15.6 &  22 \\
   10 & Sidney Crosby & C & 0.818 & $-$0.052 & 0.766 & 0.212 & 1059 & 1.04 & 13.5 &  18 \\
   12 & Tim Connolly & C & 0.501 & 0.244 & 0.745 & 0.247 & 710 & 0.90 & 8.8 &  11 \\
   15 & Alex Ovechkin & LW & 0.723 & 0.010 & 0.733 & 0.254 & 1262 & 1.42 & 15.4 &  30 \\
   16 & Jason Pominville & RW & 0.309 & 0.424 & 0.733 & 0.249 & 1052 & 0.55 & 12.9 &  10 \\
   17 & Alex Burrows & LW & 0.459 & 0.272 & 0.730 & 0.210 & 1023 & 0.94 & 12.5 &  16 \\
   \bottomrule
\end{tabular}
}
\end{center}
\end{table}

We now discuss the estimates for forwards and defensemen separately, starting with the top forwards in $APM/60$ given in Table \ref{apm60f}.  Burrows' case is an interesting one.  He did not appear in the top 10 lists for $OPM/60$ or $DPM/60$, but makes the top $APM/60$ list for forwards in Table \ref{apm60f} with balanced offensive and defensive estimates.  He has been playing frequently with the Sedin twins this year, so one might think his rating would be difficult to separate from the twins' estimates.  However, on average over the last three years, the Sedins are not among Burrows' three most frequent linemates, so his high estimates can not be attributed to statistical noise caused by frequently playing with the twins.  Burrows actually has the lowest errors in this list of players, probably because of the varied linemates that he has had over the past three years.

\begin{table}[h!]
\begin{center}
\caption{
Top 10 Defensemen in APM60 (minimum 700 minutes)
}
\label{apm60d}
{\small
\begin{tabular}{llrrrrrrrrr}
  \addlinespace[.3em] \toprule
Rk & Player & Pos & OPM60 & DPM60 & APM60 & Err & Mins & NG60 & APM & NG \\
  \midrule
 58 & Mike Green & D & 0.357 & 0.195 & 0.552 & 0.208 & 1334 & 1.02 & 12.3 &  22 \\
   61 & Nicklas Lidstrom & D & 0.242 & 0.307 & 0.549 & 0.266 & 1374 & 1.09 & 12.6 &  25 \\
   67 & Duncan Keith & D & 0.245 & 0.284 & 0.529 & 0.235 & 1532 & 0.73 & 13.5 &  19 \\
   83 & Paul Martin & D & $-$0.026 & 0.526 & 0.500 & 0.239 & 916 & 0.92 & 7.6 &  14 \\
   91 & Kent Huskins & D & 0.278 & 0.198 & 0.476 & 0.210 & 915 & 0.72 & 7.3 &  11 \\
  105 & Johnny Oduya & D & 0.342 & 0.116 & 0.458 & 0.204 & 1209 & 0.78 & 9.2 &  16 \\
  106 & Jeff Schultz & D & 0.234 & 0.220 & 0.454 & 0.219 & 1094 & 1.15 & 8.3 &  21 \\
  117 & Jan Hejda & D & 0.061 & 0.373 & 0.434 & 0.222 & 1269 & 0.08 & 9.2 &   2 \\
  120 & S. Robidas & D & 0.313 & 0.115 & 0.428 & 0.210 & 1316 & 0.12 & 9.4 &   3 \\
  142 & Andrei Markov & D & 0.370 & 0.036 & 0.405 & 0.235 & 1114 & 0.28 & 7.5 &   5 \\
   \bottomrule
\end{tabular}
}
\end{center}
\end{table}

Huskins and Schultz sneak onto the list of top defensemen in $APM/60$ in Table \ref{apm60d} with balanced ratings, despite being held off of the $OPM/60$ and $DPM/60$ top 10 lists.  We note that excluding goalies, Schultz's most common linemates are Mike Green, Alexander Ovechkin, Nicklas Backstrom, and Alexander Semin.  Schultz has accumulated fairly low goal and assist totals over the past three seasons while playing with some of league's best offensive players, and yet his $OPM/60$ estimates are still high.  In his case, the model may not be properly separating his offensive contribution from those of his teammates.  It is also possible that Schultz does a lot of little things on the ice that do not appear in box score statistics, but that contribute to his team's offensive success nonetheless.  In hockey, it is difficult to separate offense and defense.  A good defensive team, which can clear the puck from the defensive zone quickly, can help its offense by increasing its time of possession.  Likewise, a team with a good puck possession offense can help its defense by simply keeping the puck away from the opposition.  Time of possession data could help in separating offense and defense, but such data is not readily available.  The model may or may not be doing a very good job of separating the two in some cases.  See \cite{corey} for a discussion by Corey Pronman about the connection between offense and defense.


    \subsection{$APM$}
    Recall that $APM$ is a measure of the total (offensive and defensive) contribution of a player at even-strength in terms of net goals over an entire season.  A hockey fan familiar with the traditional plus-minus statistic can think of $APM$ in the same way, remembering that $APM$ has been adjusted for both the strength of a player's teammates and the strength of his opponents.
\begin{table}[h!]
\begin{center}
\caption{
Top 10 Players in APM
}
\label{apmp}
{\small
\begin{tabular}{llrrrrrrrrr}
  \addlinespace[.3em] \toprule
Rk & Player & Pos & OPM & DPM & APM & Err & Mins & NG60 & APM60 & NG \\
  \midrule
  1 & Pekka Rinne & G & NA & 23.7 & 23.7 & 6.5 & 1680 & NA & 0.845 & NA \\
    2 & Henrik Lundqvist & G & NA & 22.0 & 22.0 & 11.8 & 3223 & NA & 0.410 & NA \\
    3 & Pavel Datsyuk & C & 15.4 & 6.2 & 21.6 & 4.9 & 1186 & 1.55 & 1.091 &  31 \\
    4 & Marty Turco & G & NA & 19.7 & 19.7 & 7.1 & 2787 & NA & 0.424 & NA \\
    5 & Dan Ellis & G & NA & 19.0 & 19.0 & 5.5 & 1509 & NA & 0.757 & NA \\
    6 & Chris Mason & G & NA & 18.3 & 18.3 & 6.7 & 2384 & NA & 0.460 & NA \\
    7 & Zach Parise & LW & 12.6 & 3.0 & 15.6 & 4.6 & 1164 & 1.20 & 0.807 &  23 \\
    8 & Joe Thornton & C & 12.0 & 3.6 & 15.6 & 4.6 & 1222 & 1.08 & 0.767 &  22 \\
    9 & Alex Ovechkin & LW & 15.2 & 0.2 & 15.4 & 5.3 & 1262 & 1.42 & 0.733 &  30 \\
   10 & Mikko Koivu & C & 6.6 & 8.1 & 14.7 & 4.2 & 1032 & 0.52 & 0.852 &   9 \\
   \bottomrule
\end{tabular}
}
\end{center}
\end{table}
Goalies dominate the top of the list of best players in $APM$ in Table \ref{apmp}, as is common with many of the advanced metrics used by hockey analysts.  This trend can also be seen in Figure \ref{apmfig}.  We reiterate again that other statistics are preferred over $APM/60$ and $APM$ for estimating the contribution of goalies.

\begin{table}[h!]
\begin{center}
\caption{
Top 10 Skaters in APM
}
\label{apms}
{\small
\begin{tabular}{llrrrrrrrrr}
  \addlinespace[.3em] \toprule
Rk & Player & Pos & OPM & DPM & APM & Err & Mins & NG60 & APM60 & NG \\
  \midrule
  3 & Pavel Datsyuk & C & 15.4 & 6.2 & 21.6 & 4.9 & 1186 & 1.55 & 1.091 &  31 \\
    7 & Zach Parise & LW & 12.6 & 3.0 & 15.6 & 4.6 & 1164 & 1.20 & 0.807 &  23 \\
    8 & Joe Thornton & C & 12.0 & 3.6 & 15.6 & 4.6 & 1222 & 1.08 & 0.767 &  22 \\
    9 & Alex Ovechkin & LW & 15.2 & 0.2 & 15.4 & 5.3 & 1262 & 1.42 & 0.733 &  30 \\
   10 & Mikko Koivu & C & 6.6 & 8.1 & 14.7 & 4.2 & 1032 & 0.52 & 0.852 &   9 \\
   11 & Marian Gaborik & RW & 10.2 & 4.3 & 14.5 & 3.2 & 853 & 1.01 & 1.018 &  15 \\
   14 & Sidney Crosby & C & 14.4 & $-$0.9 & 13.5 & 3.7 & 1059 & 1.04 & 0.766 &  18 \\
   15 & Duncan Keith & D & 6.2 & 7.3 & 13.5 & 6.0 & 1532 & 0.73 & 0.529 &  19 \\
   16 & Jason Pominville & RW & 5.4 & 7.4 & 12.9 & 4.4 & 1052 & 0.55 & 0.733 &  10 \\
   17 & Nicklas Lidstrom & D & 5.5 & 7.0 & 12.6 & 6.1 & 1374 & 1.09 & 0.549 &  25 \\
   \bottomrule
\end{tabular}
}
\end{center}
\end{table}
Pavel Datysuk has won the Selke Trophy in the 2007-08 and 2008-09 seasons, and is widely regarded as one of the top two-way players in the game, at least among forwards.  He is third among players in $APM$, and he leads the list of top skaters in Table \ref{apms} by a wide margin.  It should also be pointed out that all of the other players on this list are still within two standard errors of the top spot.  Interestingly, Crosby and Ovechkin are the players with the lowest defensive estimates on this list, which hurts their overall ratings.

Since forwards dominated the list in Table \ref{apms}, we list the top 10 defensemen separately in Table \ref{apmd}.
\begin{table}[h!]
\begin{center}
\caption{
Top 10 Defensemen in APM
}
\label{apmd}
{\small
\begin{tabular}{llrrrrrrrrr}
  \addlinespace[.3em] \toprule
Rk & Player & Pos & OPM & DPM & APM & Err & Mins & NG60 & APM60 & NG \\
  \midrule
 15 & Duncan Keith & D & 6.2 & 7.3 & 13.5 & 6.0 & 1532 & 0.73 & 0.529 &  19 \\
   17 & Nicklas Lidstrom & D & 5.5 & 7.0 & 12.6 & 6.1 & 1374 & 1.09 & 0.549 &  25 \\
   20 & Mike Green & D & 7.9 & 4.3 & 12.3 & 4.6 & 1334 & 1.02 & 0.552 &  22 \\
   40 & Stephane Robidas & D & 6.9 & 2.5 & 9.4 & 4.6 & 1316 & 0.12 & 0.428 &   3 \\
   43 & Johnny Oduya & D & 6.9 & 2.3 & 9.2 & 4.1 & 1209 & 0.78 & 0.458 &  16 \\
   45 & Jan Hejda & D & 1.3 & 7.9 & 9.2 & 4.7 & 1269 & 0.08 & 0.434 &   2 \\
   50 & Zdeno Chara & D & 6.6 & 2.2 & 8.8 & 5.6 & 1441 & 0.68 & 0.367 &  16 \\
   63 & Jeff Schultz & D & 4.3 & 4.0 & 8.3 & 4.0 & 1094 & 1.15 & 0.454 &  21 \\
   64 & Keith Ballard & D & 3.8 & 4.5 & 8.3 & 4.3 & 1392 & 0.22 & 0.359 &   5 \\
   70 & Ian White & D & 6.9 & 0.6 & 7.6 & 4.1 & 1343 & 0.18 & 0.338 &   4 \\
   \bottomrule
\end{tabular}
}
\end{center}
\end{table}
The top 3 in $DPM/60$ (Table \ref{dpm60d}) are once again in the top 3 here, though in a different order.  One player we have not discussed is Jan Hejda, who seems to be an underrated player according to his $DPM$ estimate.  Looking at his traditional box score statistics, we see that during both the 2007-2008 season (+20, 13 more than the second highest Blue Jacket) and the 2008-2009 season (+23, 11 more than the second highest Blue Jacket) seasons, Hejda led his team in plus-minus by a wide margin.  His numbers were much worse during his injury shortened 2009-2010 season in which the entire Blue Jackets team struggled defensively, but his performance in the previous two seasons is one reason the model could be giving him a high estimate for $DPM$.  Also, his two most common linemates, Mike Commodore (2.71) and Rick Nash (2.69) have a higher $GA/60$ than does Hejda (2.24), which may be helping his defensive rating.

\section{Discussion of the Model}\label{discussion}  We now discuss several aspects of the formation and analysis of our model.  In Section \ref{adv} we summarize some advantages of $APM$, including that $APM$ is independent of teammates, opponents, and box score statistics.  We discuss disadvantages of $APM$ in Section \ref{disadv}, including statistical noise and difficulties in computing the estimates.  We discuss the selection of the explanatory variables and response variables in Section \ref{variables}, and selection of the observations in Section \ref{observations}.  Closely related to the selection of the components in the model are the assumptions that we made, and we discuss those in Section \ref{assumptions}.  The main assumptions discussed in that section are that in hockey teams play offense and defense concurrently, that goalies do not contribute on offense, and that there are no interactions between players.  One of the main disadvantages of $APM$ listed in Section \ref{disadv} is the errors associated with the estimates, and we discuss these errors in greater detail in Section \ref{errors}.  Also in that section, we give top 10 lists of the players with the highest and lowest errors in $APM/60$ and $APM$.  Finally, in Section \ref{futurework}, we finish with some concluding remarks and give two ideas for future work: modeling special teams situations, and accounting for the zone in which each shift starts.

\subsection{Advantages of $APM$}\label{adv}
    As with any metric of its kind, $APM$ has its advantages and disadvantages.
    As we have mentioned previously, the most important benefit of $APM$ is that a player's $APM$ does not depend on the strength of that player's teammates or opponents.  A major downside of the traditional plus-minus statistic is that it \textit{does} depend on both teammates and opponents, so it is not always a good measure of a player's individual contribution.  For example, a player on a below-average team could have a traditional plus-minus that is lower than average simply because of the linemates he plays with on a regular basis.  An average player on a hypothetical line with Wayne Gretzky and Mario Lemieux would probably have a traditional plus-minus that is very high, but that statistic would not necessarily be a good measure of his contribution to his team.  On the other hand, the coefficients in our model, which we use to estimate a player's $APM$, are a measure of the contribution of a player when he is on the ice versus when he is off the ice, independent of all other players on the ice.

    Another benefit of $APM$ is that the estimate, in theory, incorporates all aspects of the game, not just those areas that happen to be measured by box score statistics.  Box score statistics do not describe everything that happens on the ice.  For example, screening the goalie on offense and maintaining good positioning on defense are two valuable skills, but they are not directly measured using box score statistics.  $APM$ is like traditional plus-minus in that it attempts to measure how a player effects the outcome on the ice in terms of goals scored by his team on offense and goals allowed by his team on defense.  A player's personal totals in goals, assists, points, hits, and blocked shots, for example, are never used in computing $APM$.  Nothing is assumed about the value of these box score statistics and how they impact a player's and a team's performance.

    Another benefit of our model is that we make minimal \textit{ad hoc} assumptions about about which positions deserve the most credit for goal scoring or goal prevention.  We do not assume, for example, that goalies or defensemen deserve more credit than forwards in goal prevention, or that forwards deserve more of the credit when a goal is scored.  From Figure \ref{opmfig}, it seems that forwards contribute more than defensemen to goal scoring, but no such assumption was made during the formation of the model.  The one assumption about position we did make in our first model in Section \ref{model1} was that goalies do not contribute on offense (see Section \ref{goalies}).

\subsection{Disadvantages of $APM$}\label{disadv}
        One main drawback of the $APM$ estimates is statistical noise.  In particular, the standard errors in the $APM$ estimates for goalies are currently high.  A priority in future research is to take measures to reduce the errors.  We discuss the errors in detail in Section \ref{errors}.
        Another drawback of $APM$ is that the estimates do not include shootouts, and do not include the value of either penalties drawn by a player or penalties taken by a player.  A team's performance in shootouts has a big impact on their place in the standings.  Shootout specialists can be very valuable to a team during the regular season, and ideally shootout performance would be accounted for in $APM$.  Penalties drawn and taken also impact the outcome of a game.  Penalties drawn by a player lead to more power plays for that player's team, which in turn leads to more goals for his team.  Likewise, penalties taken by a player lead to more power plays for, and more goals for, the opposing team.  If the value of shootout performance, penalties drawn, and penalties taken were estimated using another method that gives results the units of goals per game or goals per season, those values could easily be combined with $APM$.

        Another difficulty with $APM$ is that the data required to calculate it is large, difficult to obtain in a usable form, and difficult to work with.  Collecting and managing the data was easily the most time-consuming aspect of this research.  Also, the data required for this model is only available (at least publicly) for very recent seasons.  This model could not be used to estimate the value of Wayne Gretzky, Mario Lemieux, and Bobby Orr, for example, independent of their teammates and opponents.

        The final downside to $APM$ is that the model requires knowledge of linear regression or linear algebra, and is not easily computed from traditional statistics. The mathematics required makes the calculation of $APM$ accessible to fewer hockey fans.  It was a priority to ensure that at least the estimates themselves could be easily understood, even if the methods of calculating them are not.

    \subsection{Selection of the variables}\label{variables}  We now make a few remarks on how we chose the explanatory variables and the response variable in the model.
    For the model, we included players who played more than 4000 shifts during the 2007-2008, 2008-2009, and 2009-2010 seasons.  In terms of minutes played, this cutoff is roughly 200 minutes per season on average.  Players with less than 4000 shifts during those seasons would have very noisy estimates which would not be very reliable.  Increasing the 4000 shift cutoff would have reduced the errors slightly, but we would have also obtained estimates for fewer players.

        In our model, the units of the coefficients are the same as the units of $y$.  We wanted to estimate a player's contribution to his team in terms of goals per 60 minutes, so we chose our response variable with those same units.  The choice of goals per 60 minutes for the units of our estimates was important because we could rate players based on this statistic, and we could also convert this rate statistic to a counting statistic, total goals over an entire season, using the minutes played by each player.  The resulting estimates have the units of goals, and they can be easily compared with traditional plus-minus as well as advanced metrics already in existence.  Also, a priority was to ensure our estimates could be easily interpreted by the average hockey fan.  Since the units of $APM$ are goals over an entire season, any hockey fan familiar with traditional plus-minus can understand the meaning of $APM$.

\subsection{Selection of the observations}\label{observations}
    Recall that we define a shift to be a period of time during the game when there are no substitutions made.  During the 2007-2008, 2008-2009, and 2009-2010 seasons, there were 990,861 shifts.  We consider only shifts that take place at even-strength (5-on-5, 4-on-4, 3-on-3), and we also require that two goalies be on the ice.  All power play and empty net situations were removed.

    We noticed some errors in the data.  There is a minimum of four players (counting goalies) and a maximum of 6 players (counting goalies) that can be on the ice for a team at the same time.  However, for some shifts, there are less than four players or more than six players on the ice for a team.  These shifts may have occurred in the middle of a line change, during which it is difficult to record in real-time which players are on or off the ice.  Such shifts were removed.  We also note that five games had missing data, and a few more games had incomplete data, such as data for just one or two periods.  The equivalent of about 10 games of data is missing out of a total of 3,690 games during the three seasons in question.  After removing shifts corresponding to empty net situations and special teams situations, and shifts where errors were identified, 798,214 shifts remained.

\subsection{Discussion of assumptions}\label{assumptions}
    Some of the assumptions used in the model require discussion.
    First, in our Ilardi-Barzilai-type model (Section \ref{model1}), we split each shift into two observations, one corresponding to the home team being on offense, and one corresponding to the away team being on offense.  We assume that in hockey a team plays offense and defense concurrently during the entire shift, and we give the two observations equal weight.  This assumption of concurrency was suggested by Alan Ryder and was used in \cite{ryder}.  In other sports, offense and defense are more distinct and more easily defined.  However, because of the chaotic nature of play in hockey, defining what it means for a team to be on offense is tricky.  Even if one could define what it means to be ``on offense", the data needed to determine if a team is on offense might not be available.

    Alteratively, we could say that the split into to two observations with equal weights was made by assuming that for each shift, a team was playing offense for half the shift and playing defense for half the shift.  One problem with this assumption is that there may be some teams that spend more time with the puck than without it.  

    \subsubsection{Goalie contribution on offense}\label{goalies}
    In our first model (Section \ref{model1}), we ultimately decided to treat goalies differently than skaters by including only a defensive variable for each goalie.  This decision was based on the the assumption that a goalie's contribution on offense is negligible.  This assumption is debatable.  There are some great puck-handling goalies, and some poor ones, and that could affect both the offensive and defensive performance of their team.  Some analysts have attempted to quantify the effects of puck handling for goalies and have come up with some interesting results.  See, for example, \cite{goalieassists}.

    While we ultimately decided against including offensive variables for goalies, we did try the model both ways, and compared the results.  We compared the offensive results for skaters, and defensive results for both skaters and goalies.  First, the defensive coefficients, the $DPM$ results, and the errors associated with them, stayed very similar for all skaters and goalies.  The offensive coefficients, and the $OPM$ estimates, stayed similar for most skaters when goalies were included, but in some extreme cases, the results varied greatly.  For example, Henrik Lundqvist's offensive rating was extremely high, with very high errors.  As a result, the offensive results for several New York Rangers were significantly lower when goalies were included.  It was as if the model was giving Lundqvist much of the credit for offensive production, while giving less credit to the skaters.  The standard errors in $OPM$ for these Rangers also increased.  Similarly, three Detroit Red Wings goalies, Dominic Hasek, Chris Osgood, and Jimmy Howard, had very low offensive ratings.  Several Detroit players saw a significant boost in offensive production when goalies were included.  Once again, the errors for these players saw a significant increase.

    One problem with these changes in offensive estimates is that the goalie ratings are extremely noisy and are not very reliable, so the effect that the goalie ratings had on the skater ratings cannot be considered reliable either.  On the other hand, there could be some positives gained from including goalies.  Recall that we do not consider empty net situations in our model, so anytime each skater is on the ice, he is on the ice with a goalie.  For teams who rely very heavily on one goalie, that goalie could get 90\% of the playing time for his team during the season.  That goalie's variable could be acting similar to an indicator variable for that team.  So the goalie's offensive estimates could be a measure of some sort of team-level effect, or coaching effect, on offensive production.  For example, a low estimate for a goalie's $OPM$ could be considered partially as an adjustment for an organizational philosophy, or a coaching system, that favors a more conservative, defensive-minded approach.

    In the end, we decided against including offensive variables for goalies in our first model because of the noisiness of the goalies' results, the effect that it had on the skaters' offensive ratings, the increase in interactions with other players, and the increase in errors that came with those changes.  Note that in our second model we do not have separate offensive and defensive variables for any of the players, including goalies, and goalies are considered on offense.  So we have one model that does not include goalies for offensive purposes, and one model that does.  When we average the results of the two models, we balance the effects of including goalies in one model, and excluding them in the other model.

   \subsubsection{Interactions between players}
    By not including interaction terms in the model, we do not account for interactions between players.  Chemistry between two particular teammates, for example, is ignored in the model.  The inclusion of interaction terms could reduce the errors.  The disadvantages of this type of regression would be that it is much more computationally intensive, and the results would be harder to interpret.

\subsection{Discussion of Errors}\label{errors}
    In the introduction, and elsewhere in this paper, we noted that Henrik and Daniel Sedin have a much higher error than other players with a similar number of shifts.  One reason for this high error could be that the twin brothers spend most of their time on the ice together.  Daniel spent 92\% of his playing time with Henrik, the highest percentage of any other player combination where both players have played over 700 minutes.  Because of this high colinearity between the twins, it is difficult to separate the individual effect that each player has on the net goals scored on the ice.  It seems as though the model is giving Henrik the bulk of the credit for the offensive contributions, and Daniel most of the credit for defense.  Henrik's defensive rating is strangely low given his low goals against while on the ice.  Likewise, Daniel's offensive rating is unusually low.

\begin{table}[h!]
\begin{center}
\caption{
Top 10 Players in Highest Err in APM60 (minimum 700 minutes)
}
\label{err60high}
{\small
\begin{tabular}{llrrrrrrrr}
  \addlinespace[.3em] \toprule
Rk & Player & Pos & APM60 & Err & Mins & Teammate.1 & min1 & Teammate.2 & min2 \\
  \midrule
 69 & Henrik Sedin & C & 0.424 & 0.328 & 1169 & D.Sedin & 83\% & R.Luongo & 76\% \\
   73 & Daniel Sedin & LW & 0.710 & 0.326 & 1057 & H.Sedin & 92\% & R.Luongo & 77\% \\
  143 & Ryan Getzlaf & C & 0.501 & 0.288 & 1116 & C.Perry & 83\% & J.Hiller & 49\% \\
  157 & B. Morrow & LW & 0.141 & 0.283 & 805 & M.Ribeiro & 73\% & M.Turco & 71\% \\
  161 & Corey Perry & RW & 0.370 & 0.282 & 1130 & R.Getzlaf & 82\% & J.Hiller & 48\% \\
  199 & T. Holmstrom & LW & 0.175 & 0.269 & 724 & P.Datsyuk & 87\% & N.Lidstrom & 51\% \\
  205 & N. Lidstrom & D & 0.549 & 0.266 & 1374 & B.Rafalski & 70\% & P.Datsyuk & 49\% \\
  210 & David Krejci & C & 0.642 & 0.265 & 912 & T.Thomas & 60\% & B.Wheeler & 49\% \\
  218 & N. Kronwall & D & 0.405 & 0.264 & 1055 & B.Stuart & 46\% & C.Osgood & 46\% \\
  221 & Jason Spezza & C & 0.390 & 0.263 & 1075 & D.Alfredss & 60\% & D.Heatley & 59\% \\
   \bottomrule
\end{tabular}
}
\end{center}
\end{table}
        The ten players with the highest error in $APM/60$ are shown in Table \ref{err60high}.  Note that if we do not impose a minutes played minimum, the list is entirely made up of players who played less than 200 minutes, so we have restricted this list to players that have played more than 700 minutes on average over the last three seasons.  The Sedins have significantly larger errors than the next players in the list, and all of the players in this list are ones who spent a large percent of their time on the ice with a particular teammate or two.

\begin{figure}[h!] \centering
            \includegraphics[width=\textwidth]{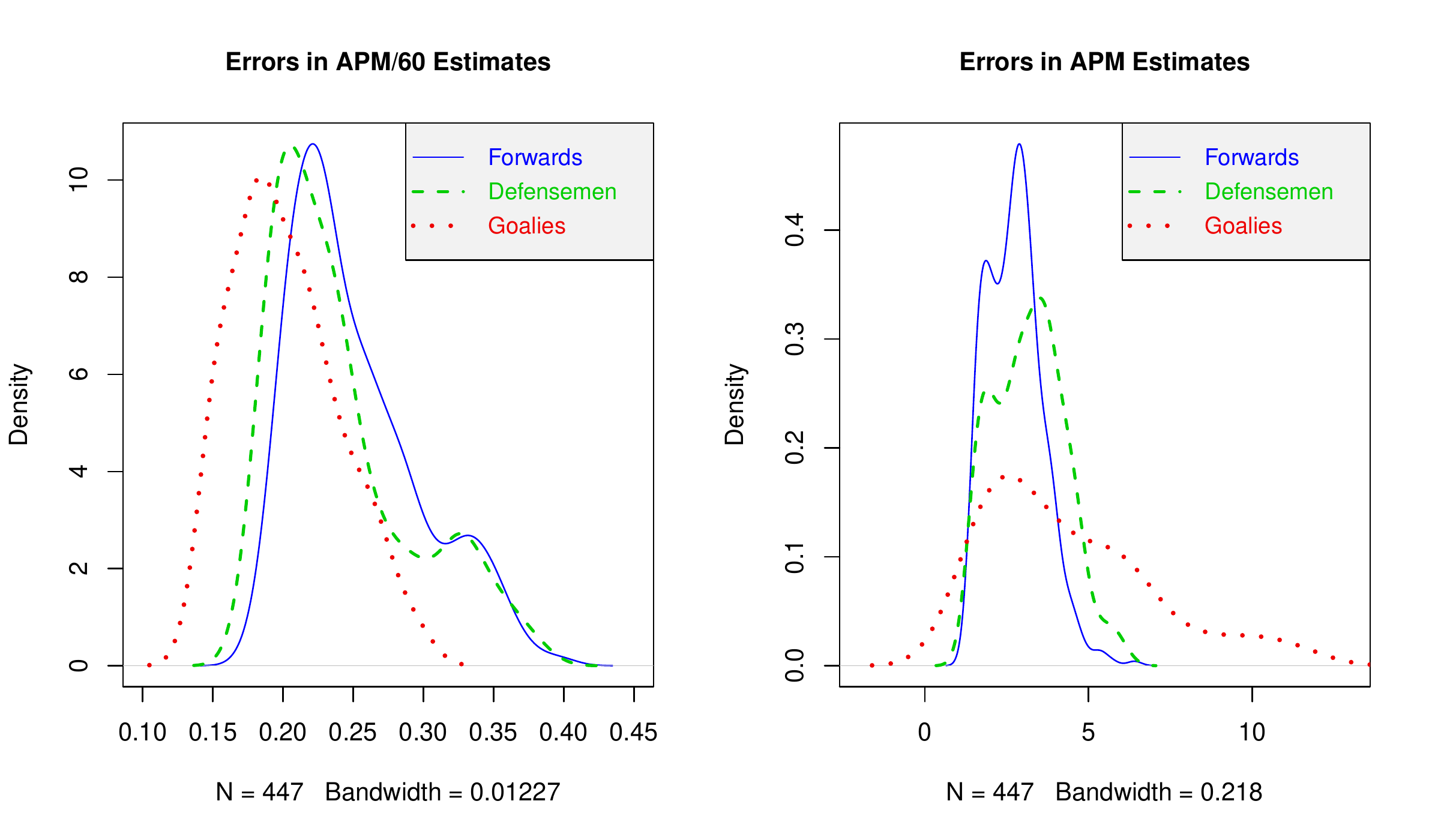}
            \caption{Kernel Density Estimation for $APM/60$ Errors and $APM$ Errors.}\label{errorsfig}
        \end{figure}

        In Table \ref{err60lowa}, we list the players with the lowest errors in $APM/60$.
\begin{table}[h!]
\begin{center}
\caption{
Top 10 Players in Lowest Err in APM60 (minimum 700 minutes)
}
\label{err60lowa}
{\small
\begin{tabular}{llrrrrrrrr}
  \addlinespace[.3em] \toprule
Rk & Player & Pos & APM60 & Err & Mins & Teammate.1 & min1 & Teammate.2 & min2 \\
  \midrule
  1 & Mike Smith & G & 0.222 & 0.144 & 1704 & M.St. Loui & 27\% & V.Lecavali & 26\% \\
    2 & D. Roloson & G & 0.141 & 0.145 & 2259 & T.Gilbert & 24\% & S.Staios & 24\% \\
    3 & Martin Biron & G & 0.012 & 0.149 & 2077 & B.Coburn & 28\% & K.Timonen & 25\% \\
    4 & J. Labarbera & G & 0.165 & 0.150 & 1205 & A.Kopitar & 22\% & P.O'Sulliv & 20\% \\
    5 & Cam Ward & G & 0.317 & 0.151 & 2655 & E.Staal & 32\% & T.Gleason & 31\% \\
    6 & Alex Auld & G & 0.245 & 0.152 & 1409 & C.Phillips & 17\% & D.Heatley & 15\% \\
    7 & A. Niittymaki & G & 0.177 & 0.154 & 1448 & B.Coburn & 19\% & K.Timonen & 17\% \\
    8 & Ilja Bryzgalov & G & 0.203 & 0.154 & 2942 & Z.Michalek & 35\% & E.Jovanovs & 32\% \\
    9 & Marty Turco & G & 0.424 & 0.154 & 2787 & S.Robidas & 35\% & T.Daley & 35\% \\
   10 & Manny Legace & G & 0.253 & 0.155 & 1680 & B.Jackman & 28\% & E.Brewer & 24\% \\
   \bottomrule
\end{tabular}
}
\end{center}
\end{table}
 Goalies dominate this list, partially because of playing time, and partially because goalies share the ice with a wider variety of players than skaters do.  Also, with the exception of Turco and Ward, all of the goalies in the list have the benefit of playing with more than one team, further diversifying the number of players that they have played with.  While goalies have lower errors in $APM/60$ than skaters do, that changes with playing-time dependent $APM$ statistic (see Figure \ref{errorsfig}).

 If we remove goalies from consideration, we get the top 10 skaters in lowest standard errors as shown in Table \ref{err60lows}.
\begin{table}[h!]
\begin{center}
\caption{
Top 10 Skaters in Lowest Err in APM60 (minimum 700 minutes)
}
\label{err60lows}
{\small
\begin{tabular}{llrrrrrrrr}
  \addlinespace[.3em] \toprule
Rk & Player & Pos & APM60 & Err & Mins & Teammate.1 & min1 & Teammate.2 & min2 \\
  \midrule
 20 & J. Bouwmeester & D & 0.170 & 0.177 & 1532 & T.Vokoun & 50\% & M.Kiprusof & 29\% \\
   24 & Olli Jokinen & C & 0.172 & 0.178 & 1165 & T.Vokoun & 30\% & M.Kiprusof & 27\% \\
   28 & D. Seidenberg & D & 0.090 & 0.179 & 1067 & C.Ward & 45\% & T.Vokoun & 28\% \\
   29 & C. Ehrhoff & D & 0.296 & 0.180 & 1285 & E.Nabokov & 54\% & R.Luongo & 28\% \\
   30 & Ian White & D & 0.338 & 0.181 & 1343 & V.Toskala & 53\% & M.Stajan & 30\% \\
   31 & Bryan Mccabe & D & 0.193 & 0.181 & 1172 & T.Vokoun & 53\% & V.Toskala & 22\% \\
   35 & P. O'Sullivan & C & 0.074 & 0.183 & 1042 & A.Kopitar & 30\% & J.Labarber & 23\% \\
   36 & Greg Zanon & D & 0.003 & 0.183 & 1307 & D.Hamhuis & 30\% & N.Backstro & 27\% \\
   39 & Keith Ballard & D & 0.359 & 0.184 & 1392 & T.Vokoun & 48\% & D.Morris & 23\% \\
   40 & Lee Stempniak & RW & 0.463 & 0.184 & 982 & M.Legace & 27\% & V.Toskala & 25\% \\
   \bottomrule
\end{tabular}
}
\end{center}
\end{table}
Most of these players are defensemen and have been on the ice for a high number of minutes.  Every player in Table \ref{err60lows} has played for two or more teams during the past three seasons.  Stempniak, who has the lowest minutes played on the list, probably made the list because he has played for three different teams.  Also, Stempniak shared the ice with his most common linemate, Manny Legace, for just 27\% of his time on the ice.

    We can look at the overall trend in $APM/60$ errors and $APM$ errors in Figure \ref{errorsfig}.
    The trends we noticed in the top 10 lists continue outside of the top 10.  In particular, it appears that goalies tend to have the lowest errors in $APM/60$.  The downside is that since many goalies get much more playing time than skaters, those goalies have much noisier $APM$ estimates.

\subsection{Future work and conclusions}\label{futurework}
    We highlight two improvements that could be made to our model.
 The most important addition to this work would probably be to include a player's offensive and defensive contributions in special teams situations.  While performance at even-strength is a good indicator of a player's offensive and defensive value, some players seem to have much more value when they are on special teams.  Teemu Selanne is an example of one player who has the reputation of being a power play goal-scoring specialist, and we could quantify his ability using an estimate that includes special teams contributions.

 Another improvement we could make is accounting for whether a shift starts in the offensive zone, defensive zone, or neutral zone, and accounting for which team has possession of the puck when the shift begins.  The likelihood that a goal is scored during a shift is dependent on the zone in which a shift begins and is dependent on which team has possession of the puck when the shift begins.  See, for example, \cite{thomaspossession}.  This fact could be affecting the estimates of some players.  Players who are relied upon for their defensive abilities, for example, may start many of their shifts in their own zone.  This trend could result in more goals against for those players than if they had started most of their shifts in their offensive zone.  In the current model, there is no adjustment for this bias.

We believe that $APM$ is a useful addition to the pool of hockey metrics already in existence.  The fact that $APM$ is independent of teammates and opponents is the main benefit of the metric.  $APM$ can be improved by addressing special teams and initial zones, and reducing the statistical noise is a priority in future research.  We hope that GM's, coaches, hockey analysts, and fans will find $APM$ a useful tool in their analysis of NHL players.

\bibliographystyle{BEPress}
\bibliography{generalbib}

\end{document}